\documentclass[aps,pre,reprint,superscriptaddress,amsmath,amssymb,floatfix]{revtex4-2}
\usepackage{graphicx}  
\usepackage{bm}        
\usepackage[dvipsnames]{xcolor} 
\usepackage{hyperref}

\hypersetup{
    colorlinks=true,      
    linkcolor=Blue,     
    citecolor=Blue,     
    urlcolor=Blue       
}
\usepackage{soul}

\newcommand{\beginsupplement}{%
  \section*{Appendix}%
  \setcounter{figure}{0}%
  \renewcommand{\thefigure}{S\arabic{figure}}%
  \renewcommand{\theHfigure}{S\arabic{figure}}%
}

\begin{document}

\title{Optimal multi-parameter control of trapped active matter}
\author{Luke K. Davis}
\email{luke.davis@ed.ac.uk} 
\affiliation{%
School of Mathematics and Maxwell Institute for Mathematical Sciences, University of Edinburgh, EH9 3FD, Scotland
}%
\affiliation{%
Higgs Centre for Theoretical Physics, University of Edinburgh, EH9 3FD, Scotland
}%

\begin{abstract}
The realization of efficient micro-machines built from active matter requires precise thermodynamic control far from equilibrium. Despite theoretical progress, the focus on single-parameter driving, coupled with strict theoretical assumptions, limits efforts to capture modern multi-parameter control experiments. Here, guided by careful theoretical considerations, we develop a transparent computational framework based on exact-gradient descent via automatic differentiation. We derive optimal protocols for a wide range of multi-parameter problems—involving trap stiffness, trap center, and particle activity—to minimize the thermodynamic work or heat. We demonstrate that smoothed, experimentally plausible protocols—obtained by assigning kinetic costs to the controls—achieve near-optimal efficiencies comparable to discontinuous ``bang-bang'' solutions. By exploring both open- and closed-loop control, we find the dynamical coupling between parameters leads to genuinely new strategies, including symmetry breaking in optimal activity cycles and non-monotonic trap stiffness controls. Further, we identify regimes where initial measurement and multi-parameter flexibility combine to improve efficiency. Finally, we reveal that the naive simultaneous execution of independently optimized controls incurs only slightly more work than the full multi-parameter solutions. Taken together, our work elucidates the non-equilibrium physics of multi-parameter control and provides robust, scalable strategies for controlling active matter.
\end{abstract}
\maketitle

\section{Introduction}

Active systems, comprised of constituents that convert latent fuel to sustain individual dynamics, display a broad range of emergent non-equilibrium behaviors, such as swarming and motility-induced phase separation (MIPS) \cite{Ramaswamy2010,Cates2015,Marchetti2013}. Such phenomenology is typically not seen at equilibrium (\emph{i.e.,} in passive systems) and has required refined, and in many cases new, statistical physics models and tools to understand it \cite{Solon_2015,Datta2022,Fodor2016,Ekeh2020,Davis2024}. Naturally, there is much scope in the control and design of active materials, devices, and engines that harness these behaviors in a controlled manner to perform useful functions \cite{DiLeonardo2010,Norton2020,Piro2021,Falk2021,Fodor2021,Shankar2022,Gupta2023,Davis2024,Casert2024,Wareham2025,Takatori2025,Alvarado2026}. The far-from-equilibrium nature of active matter has made precise and energy-efficient control difficult, mainly due to the challenge of accounting for both the active dissipation and the dissipation resulting from external manipulation \cite{Piro2022,Zhong2022,Baldovin2023,Davis2024,Soriani2025,Olsen2025,GarciaMillan2025,Cocconi2024,Cocconi2025,Schuttler2025,Zhong2025,Baldovin2026}. 

Theoretical progress in this direction has revealed, among other things: (i) that finite-time, as opposed to infinitely slow, control protocols minimize dissipation \cite{Davis2024,Soriani2025}, (ii) that closed-loop (feedback) controls differ greatly from passive closed-loop controls \cite{GarciaMillan2025,Schuttler2025}, and (iii) that work may be extracted by leveraging active dynamics \cite{Pietzonka2019a,Pietzonka2019b,Fodor2021,Cocconi2023}. Despite this progress, which has mainly focused on single-parameter control of one-body systems and in the asymptotic regimes of slow \cite{Kamizaki2022,Zhong2022,Davis2024,Soriani2025,Baldovin2026} (or fast \cite{Blaber2021}) driving, the control scenarios treated in current theory lag behind the complexity of control that is possible in experiments \cite{Truong2026,Alvarado2026}. Encouragingly, for the driving of a trapped (passive) Brownian particle, previous work has shown that optimal multi-parameter controls result in some reduction in thermodynamic cost, as compared with lower dimensional control \cite{Blaber2022}. This gap in active control complexity currently thwarts the realization of efficient control for a range of active matter systems.
\begin{figure*}[t!]
    \centering
    \includegraphics[width=0.99\linewidth]{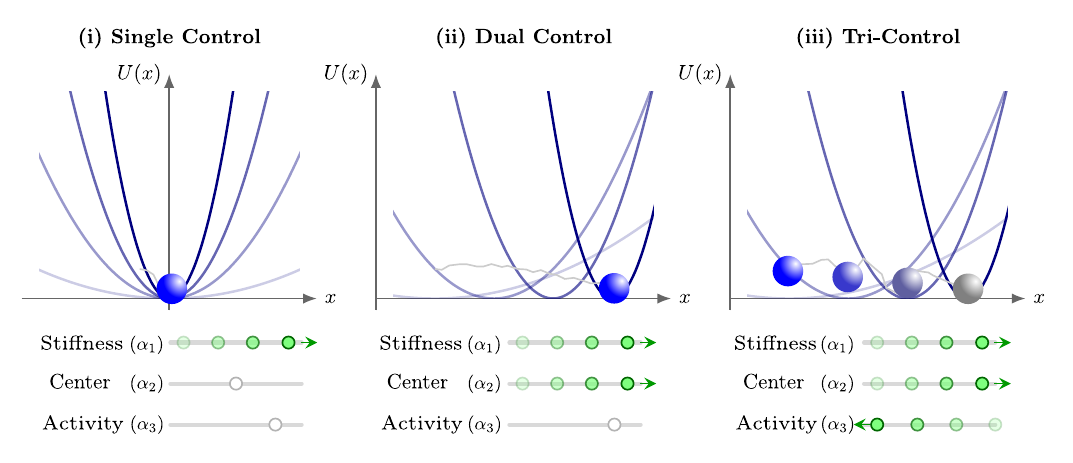}
    \caption{\textbf{Examples of the control scenarios explored in this paper}. \textbf{(i)} The single-control problem, where only one of the controls changes as a function of time, with the strengthening of the stiffness shown as an example. \textbf{(ii)} The dual-control problem, where two controls are subject to time-dependent protocols, with the other control fixed. \textbf{(iii)} The tri-control scenario, with dual control of the trap coupled to a time-dependent activity protocol.}
    \label{fig:Sketch}
\end{figure*}

Computational and numerical methods are natural approaches to address the growing gap between experiments and analytical theory. Indeed, this is starting to appear in the area of optimal non-equilibrium control \cite{Yan2022,Chennakesavalu2023,Engel2023,Casert2024}. This includes the use of numerical simulations to implement the model and machine-learning or artificial-intelligence approaches to perform the optimization of controls \cite{Falk2021,Yan2022,Nasiri2022,Casert2024}. The use of gradient descent via automatic differentiation (AD)--which obtains exact gradients by automatically propagating the chain rule through computer programs--shows promise in terms of both the transparency of the method and its ability to accurately solve complex optimization problems in statistical physics \cite{Goodrich2021,Engel2023}. Despite its previous successes, such AD approaches have yet to be fully leveraged in the optimal control of active matter.

Here, using exact-gradient AD methods, we solve various optimal multi-parameter control problems for a single active particle in a one-dimensional trap (see Fig.~\ref{fig:Sketch}). The deliberate choice of a minimal physical model reflects our ambition to systematically explore a wide range of novel multi-parameter active-control scenarios, and to extract principles that should form the basis for more involved systems. Moreover, keeping the physics simple will facilitate, and encourage, empirical testing of the resulting controls. First, we verify our method by reproducing analytical optimal protocols for the single-parameter control of both passive and active particles. We also compare our method to optimal protocols obtained by neural computation \cite{Casert2024}. Going beyond what has currently been explored theoretically and computationally, we solve for optimal protocols involving up to three simultaneous control parameters: the trap stiffness, trap location, and the degree of activity.

The remainder of the paper is organized as follows. In section \ref{sec:1} we introduce the general control set-up. In section \ref{sec:2} we define the thermodynamic cost functions, discuss their mathematical structures, and motivate the regularization used to obtain smooth protocols. Section \ref{sec:3} outlines the exact-gradient optimization framework based on automatic differentiation, and section \ref{sec:4} introduces the trapped active Ornstein-Uhlenbeck particle model. In section \ref{sec:5} we present the main results, beginning with benchmark single-control problems and then moving to dual- and tri-control scenarios in both open- and closed-loop settings. We conclude in section \ref{sec:6} with a discussion of the physical implications of our results and possible future directions.

\section{Control set-up and notation}
\label{sec:1}

We consider finite-time controls in the form of time-dependent model parameters $\underline{\alpha}(t) = (\alpha_1(t),\ldots,\alpha_n(t))^\intercal$ with end-points $\underline{\alpha}(t\leq 0^-)=\underline{\alpha}_0$ and $\underline{\alpha}( t_p^+ \leq t)=\underline{\alpha}_{f}$, where $t_p$ is the protocol duration. The state is described by some vector $\underline{X}(t)$. Mathematically, the controls obey a master protocol function $F$:
\begin{equation}
\begin{aligned}
     F(t\leq 0^-,\underline{\alpha}_0) &= \underline{\alpha}_0, \\
     F(t''>t', F(t'>t,\underline{\alpha}_0)) &= F(t''+t',\underline{\alpha}_0), \\
     F(t''\geq t_p^+,\underline{\alpha}_0) &= \underline{\alpha}_f.
\end{aligned}
\end{equation}

Here, optimal control involves minimizing a cost $J$ which takes the general form:
\begin{equation}
\begin{aligned}
        J[F] &:= B(F_0, F_f, \underline{X}_0, \underline{X}_f) + \int_0^{t_p} \mathcal{L}(F(t), \underline{X}(t)) dt,
\end{aligned}  
\end{equation}
where $B(\ldots)$ is a boundary term (terminal cost) and $\mathcal{L}(\ldots)$ is the running cost, which is in general sensitive to the time-dependent forms of the protocols. Later, depending on the choice of thermodynamic cost, constraints, and the type of control, we specify $B$ and $\mathcal{L}$.

We explore both open-loop (OL, or feed-forward) and closed-loop (CL, or feedback) controls \cite{Bechhoefer2021}. In open-loop control one executes the protocols for the specified time-window in the absence of any measurement, whereas in closed-loop control one incorporates measurement of the system to inform control.

In practice, the control is subject to noise, through stochasticity in the system dynamics, and/or measurements of quantities when considering closed-loop control. The word ``optimal'' is therefore context dependent, with the optimal master protocol $F^\ast$ taken to mean one of the following: $J [F^\ast] \leq J [F]$ (``instance-wise optimal''), $\langle J [F^\ast] \rangle \leq \langle J [F] \rangle$ (``noise-averaged optimal''), and $\langle J [F^\ast] \vert_{x_0}\rangle \leq \langle J [F]\vert_{x_0} \rangle$ (``conditioned noise-averaged optimal''). Instance-wise optimal means there is no $F$ that can beat $F^\ast$ for a fixed noise trajectory and prescribed initial condition. Noise-averaged optimal means $F^\ast$ is uniquely optimal for an average over realizations of the noise and initial conditions. Conditioned (noise-averaged) optimal means $F^\ast$ is uniquely optimal for the system conditioned on an initial measurement of some observable $x(t=0) = x_0$, and then averaged over realizations of the noise.

\section{Thermodynamic costs}
\label{sec:2}

We explore minimization of the coarse-grained heat dissipated to the bath (Sekimoto relation) given as:
\begin{equation}
\label{eq:SekimotoHeat}
\langle J_\text{H} \rangle = \frac{1}{\mu}\int_0^{t_p}\! dt\,\left\langle \dot r \circ \big(\dot r - \sqrt{2D}\,W\big)\right\rangle,
\end{equation}
where $\circ$ is the Stratonovich product \cite{Sekimoto1998,Seifert2012,Fodor2016,Davis2024,Soriani2025}. We also explore minimization of the work to the controller \cite{Crooks1998,Schmiedl2007,Seifert2012}:
\begin{equation}
\label{eq:Work}
 \langle J_\text{W} \rangle = \int_0^{t_p} dt~\underline{\dot{\alpha}}^\intercal(t) \cdot \Big\langle \underline{\partial}_{\alpha} \phi(r(t),\underline{\alpha}(t)) \Big\rangle.
\end{equation}
The above thermodynamic relations grant a decomposition into a boundary term plus an ``action'' term,
$
\langle J \rangle = B + \int_0^{t_p} \mathcal{L}(t) \, dt,
$
with the work cost \eqref{eq:Work} being decomposed as
\begin{equation}
\begin{aligned}
\label{eq:work_terms}
B_\text{W}=0, \quad & \mathcal{L}_\text{W}(t)=\dot{\underline{\alpha}}^\intercal(t) \cdot \underline{\lambda}(t), \\
\underline{\lambda}(t):=\Big\langle & \underline{\partial}_{\alpha}\phi\big(r(t),\underline{\alpha}(t)\big)\Big\rangle.
\end{aligned}
\end{equation}
Likewise, substituting the dynamics $\dot r - \sqrt{2D}\,W = -\mu\,\partial_r\phi + v$ into Eq.~\eqref{eq:SekimotoHeat} and applying the Stratonovich chain rule $d\phi = \partial_r\phi \circ dr + \underline{\partial}_{\alpha}\phi \cdot d\underline{\alpha}$ yields the heat cost:
\begin{equation}
\label{eq:heat_decomposition}
\begin{split}
\langle J_\text{H} \rangle &= B_\text{H} + \int_0^{t_p} \mathcal{L}_\text{H}(t) \, dt, \\
B_\text{H} &= \Big\langle \phi\big(r(0),\underline{\alpha}(0)\big) - \phi\big(r(t_p),\underline{\alpha}(t_p)\big)\Big\rangle, \\
\mathcal{L}_\text{H}(t) &= \dot{\underline{\alpha}}^\intercal(t) \cdot \underline{\lambda}(t) + \frac{1}{\mu} \Big\langle \dot r(t) \circ v(t) \Big\rangle,
\end{split}
\end{equation}
with this decomposition expressing the first law of thermodynamics.

Thus, both thermodynamic objectives contain a term affine (linear) in $\dot{\underline{\alpha}}$ with coefficient $\underline{\lambda}$. If, for multi-parameter control, $\underline{\lambda}(t)$ were a single-valued function of the instantaneous point $\underline{\alpha}(t)$ (such that $\lambda = \lambda_a(\underline{\alpha}) \, d\alpha_a$ defined a 1-form on control space), then for cyclic protocols one could appeal to Stokes' theorem through the associated control-space ``Poisson bracket'' (curl) $
\{\lambda_i, \lambda_j\}_{\underline{\alpha}} := \partial_{\alpha_i} \lambda_j(\underline{\alpha}) - \partial_{\alpha_j} \lambda_i(\underline{\alpha})$.
Generically, however, in finite-time driving $\underline{\lambda}(t) = \langle \underline{\partial}_{\alpha}\phi(r(t),\underline{\alpha}(t)) \rangle$ depends on the non-stationary state (and hence on the protocol history); it does not define a 1-form on $\underline{\alpha}$-space, and the Stokes reduction is essentially blocked. Moreover, even when an instantaneous map $\underline{\lambda}(\underline{\alpha})$ exists, active non-equilibrium steady states will typically exhibit, for $n >1$,
\begin{equation}
\label{eq:control_curl}
 \partial_{\alpha_i} \lambda_j(\underline{\alpha}) - \partial_{\alpha_j} \lambda_i(\underline{\alpha}) \neq 0.
\end{equation}
implying path-dependent cyclic costs. Accounting for this path-dependency makes analytical solution of optimal multi-parameter control problems difficult, and would also apply to multi-parameter generalizations of response-based control frameworks (\emph{e.g,} \cite{Davis2024,Zhong2025,Soriani2025}), requiring more careful treatment in obtaining optimal protocols. 

\textit{Affine structure results in protocol jumps and numerical chattering}:-Using Pontryagin's maximum principle (PMP) one can show, under quite general conditions, that when the cost function is affine in the controls (linear in $\underline{\dot{\alpha}}$) jumps in the protocols at the end-points, or bang-bang protocols, result \cite{Athans1966,Liberzon2012,Corella2024}. Then, given this general mathematical fact it is to no surprise that the thermodynamic costs, dissipation and work affine in control speeds, have been shown to exhibit jumps irrespective of the physical properties of the system \cite{Schmiedl2007,Schuttler2025}. Further, frameworks which impose smooth and continuous conditions on the protocols obtain, as direct output, cost functions which contain non-affine terms $\propto \underline{\dot{\alpha}}^q$, with $q=2m$, $m=\mathbb{Z}$, and typically $q=2$ \cite{Davis2024}.

To see this more clearly, we give a simple scaling argument which shows why discontinuities are not penalized by the affine terms in the dissipation and work. Consider varying a single component $\alpha_a$ by an amount $\Delta\alpha_a$ over a short time window of width $\varepsilon$, so that $\dot{\alpha}_a \sim \Delta\alpha_a / \varepsilon$. The affine contribution satisfies:
\begin{equation}
\label{eq:affine_scaling_jump}
\begin{aligned}
    \int_{0}^{\varepsilon} dt \, \lambda_a(t) \dot{\alpha}_a(t) &= \int_{\alpha_a(0)}^{\alpha_a(\varepsilon)} \lambda_a \big( t(\alpha_a) \big) \, d\alpha_a \underset{\varepsilon \rightarrow 0}{=} \bar{\lambda}_a \Delta\alpha_a,
\end{aligned}
\end{equation}
which remains finite and (to leading order) independent of $\varepsilon$. Hence, unlike a quadratic regularization $\int dt \, \dot{\alpha}_a^2 \sim (\Delta\alpha_a)^2 / \varepsilon$ that diverges as $\varepsilon \to 0$, the thermodynamic objectives place no direct cost on arbitrarily fast changes: minimizing sequences can squeeze changes in $\underline{\alpha}$ into vanishingly short boundary layers, producing end-point jumps (see Appendix \ref{app:Jumps} for more details).

When numerically solving for optimal controls, where time must be discretized, the affine cost structure and PMP often result in infinitely many control oscillations in a finite-time that, in practice, results in chattering between protocol end points; this is also known in the optimal control literature as Fuller's phenomenon \cite{Fuller1963,Kupka1990} (see also Appendix \ref{app:Chattering}). These protocol oscillations naturally present problems for numerical approaches as such approaches typically only work well with piecewise continuous protocol functions \cite{Zhu2016,Robin2022}.

To progress in the search for efficient multi-parameter controls, whilst avoiding the non-trivial chattering issues, one can resort to adding a regularization term to the thermodynamic cost that essentially introduces a kinetic cost to the control ``dials''. Another, more physical, reason to add kinetic costs to the controls comes from experimental limitations, where instantaneous (faster than light) protocol switching is unavailable. The simplest such term is referred to as Tikhonov regularization which is of the lowest order (quadratic in control speeds). Then, the \textit{regularized} thermodynamic costs take the following form:
\begin{equation}
\begin{aligned}
\label{eq:JRCost}
        \langle \tilde{J}_{H,W} \rangle &= \langle J_{H,W} \rangle + J_R, \\
        &= \langle J_{H,W} \rangle + \frac{m_\varepsilon}{2} \int_0^{t_p}  \underline{\dot{\alpha}}^\intercal \underline{\dot{\alpha}}~dt,
\end{aligned}
\end{equation}
where $J_{H,W}$ means $J_{H}$ or $J_{W}$, and where $m_\varepsilon>0$ is typically a small control ``mass'', and the last term in the above is the kinetic control cost. We stress that, physically, the term $J_R$ is a rudimentary model of the energetic costs in turning the control ``dials'', and, mathematically, one can view it purely as a regularization to circumvent Fuller's phenomenon.

Thus, even for small $m_\varepsilon$, protocol jumps--or otherwise very steep gradients in the controls--will be penalized, and so any trustworthy optimization process would guide protocols (now in the context of the regularized costs) to take piece-wise continuous shapes. Solely based on the fact that the family of piece-wise continuous controls is a subset of the family of controls with admissible jumps one has:
\begin{equation}
\label{eq:RegGTRExact}
     \langle \tilde{J}_{H,W}[\tilde{F}^\ast] \rangle \geq \langle {J}_{H,W}[\tilde{F}^\ast] \rangle,
\end{equation}
with equality strictly occurring in the $m_\varepsilon \rightarrow 0$ limit. Thus, already we have compromised true optimality for regularity, though as we show in some test cases the increase in the true thermodynamic cost \emph{i.e.,} $\langle \tilde{J}_{H,W} \rangle - J_R$, is only marginal.

\textit{State-to-state transformations:-}At the end of the protocol one is free to (or to not) enforce a steady-state, or state-to-state (STS) transformation, through the cost function \cite{Baldovin2023}. Focusing on the heat, the simplest way to impose a STS transformation is through the boundary term $B_H$:
\begin{equation}
\begin{aligned}
        B_H =& \big\langle \phi\big(r(0),\underline{\alpha}(0)\big) \big\rangle_0 - \big\langle \phi\big(r(t_p),\underline{\alpha}(t_p)\big)\big\rangle, \\
        \underset{\text{sts}}{=}& \big\langle \phi\big(r(0),\underline{\alpha}(0)\big) \big\rangle_0 - \big\langle \phi\big(r(t_p+\alpha'_3),\underline{\alpha}(t_p)\big)\big\rangle_0, \\
\end{aligned}
\end{equation}
where $\langle \ldots \rangle_0$ denotes a steady state average and $\alpha'_3$ is the slowest relaxation time of the system. This has also been explored by changing the upper-limit of the time-integral from $t_p$ to $t_p + \alpha'_3$ \cite{Soriani2025}, and essentially results in the same physical constraint.

\section{Exact gradient descent}
\label{sec:3}

To obtain the optimal non-equilibrium protocols, we first parameterize each time-dependent control parameter as a piecewise-linear function defined by $M$ uniformly spaced nodes across the protocol duration $[0, t_p]$. We then optimize through gradient descent, which typically involves the following:
$
\underline{\alpha}^{(n+1)} = \underline{\alpha}^{(n)} - \eta \nabla_{\underline{\alpha}}\langle J_{W,H} \rangle (\underline{\alpha}^{(n)})$,
with learning rate $\eta>0$. However, for complex optimization problems, this often results in poor convergence and so in practice we employ the \texttt{Adam} optimizer, which maintains exponentially moving averages of the first and second moments of the gradient \cite{Kingma2014} (see Appendix \ref{app:GDAD}). Rather than relying on finite-difference approximations or derivative-free evolutionary algorithms, we leverage reverse-mode automatic differentiation (AD) using the \texttt{JAX} library \cite{jax2018github}. AD allows us to efficiently and precisely backpropagate the chain rule through the entire finite-time trajectory. This yields the exact analytical gradients of the thermodynamic costs with respect to all control nodes simultaneously. These exact gradients are then passed to a momentum-based gradient descent algorithm (\texttt{Adam} \cite{Kingma2014}) to iteratively update the protocols until the (regularized) thermodynamic cost converges to a minimum. By directly differentiating the exact physical dynamics, this approach provides a transparent, deterministic, and highly scalable computational framework for solving complex control problems. Full mathematical and algorithmic details of the protocol parameterization and the optimization are provided in Appendix \ref{app:GDAD}.

\begin{figure}[b!]
    \centering
    \includegraphics[width=0.8\linewidth]{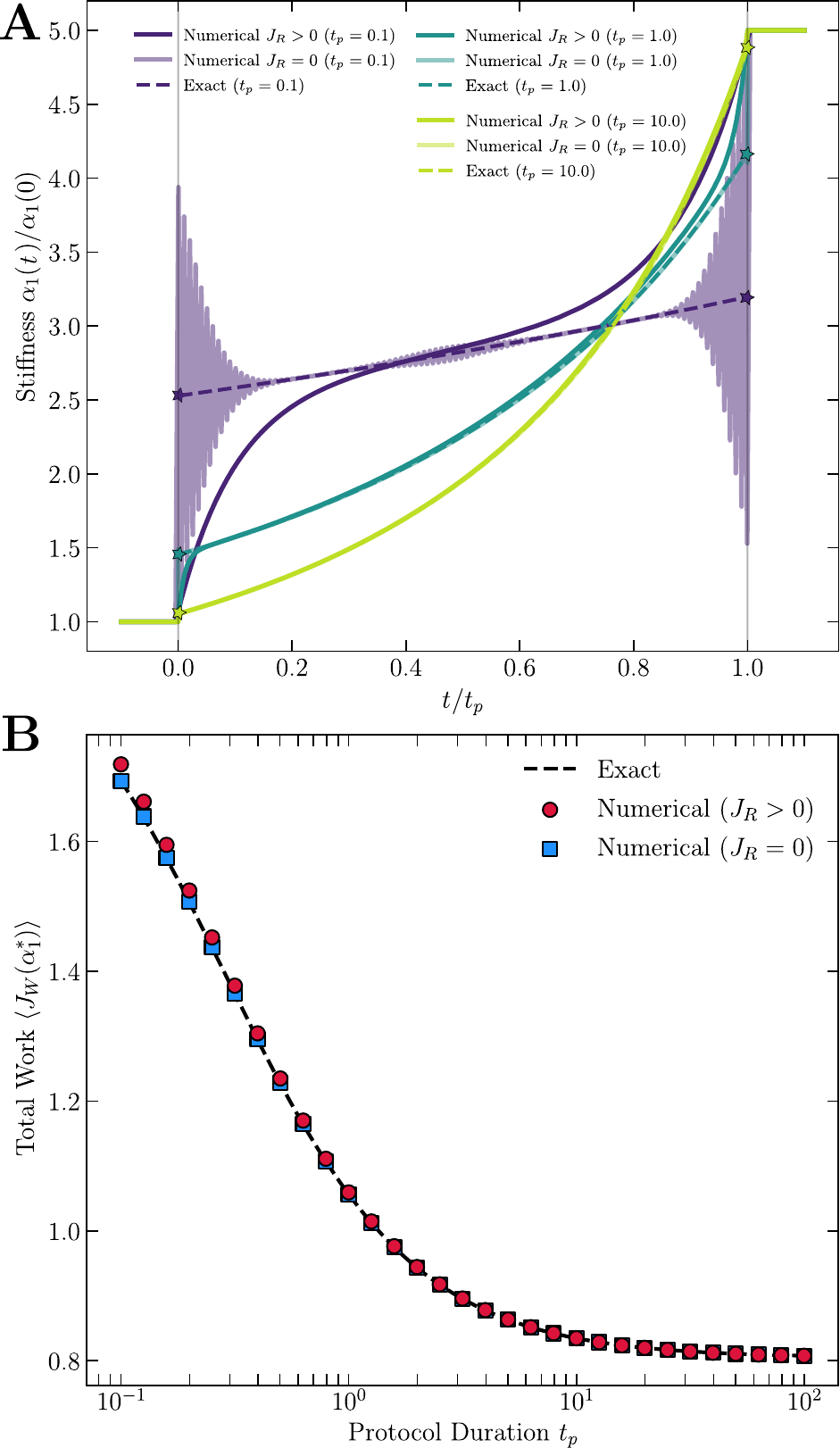}
    \caption{\textbf{Comparing our method to the exact results of Schmiedl and Seifert (2007)} \cite{Schmiedl2007}. \textbf{A} Work minimizing single parameter control of the trap stiffness as a function of the normalized protocol time $t/t_p$. Two cases of numerical optimal protocols are shown: one accounting a kinetic control cost $J_R > 0$ (regularized, \eqref{eq:JRCost}) and the other have no cost on control speed $J_R = 0$. Due to the discrete-time nature of the protocol, chattering occurs in numerical optimal protocols found using the unregularized cost function. The stars denote inner ($0^+\leq t \leq t_p^-$) discontinuities in the protocol. \textbf{B} Total work for the predicted optimal stiffness protocols as a function of protocol duration. from Model parameters: $\alpha_1(t=0^-) = 1$, $\alpha_1(t=t_p^+) = 5$, $\alpha_1(t) \in \mathbb{R}$, $D=1$ and $\mu=1$. Optimization parameters: $M=500$, $m_\varepsilon = 10^{-4}$.}
    \label{fig:schmeidltest}
\end{figure}

\section{Model}
\label{sec:4}

The focus here is to explore more complex, multi-parameter, control scenarios and in so doing we keep to the simplest model consisting of a single trapped active Ornstein-Uhlenbeck (OU) particle. The particle's position $r \in \mathbb{R}$ undergoes the following overdamped dynamics:
\begin{align}
\dot r &= -\mu\,\partial_r \phi(r) + \sqrt{2D}\,W + v,
\\
\phi &= \frac{\alpha_1}{2}\,(r-\alpha_2)^2,
\\
\alpha_3\,\dot v &= -v + \sqrt{2D'}\,W',
\end{align}
where $ \mu $ is the mobility, $ D=\mu k_B T $ is the passive diffusion coefficient, $\alpha_1$ is the trap strength, $\alpha_2$ the trap location,  $v$ is the self-propulsion obeying an OU process with stationary statistics $ \langle v \rangle = 0 $, $ \langle v^2 \rangle = D'/\alpha_3 $, $\alpha_3$ is the persistence time of the OU process, and $ W, W' $ are independent unit-variance Gaussian white noises.

In this model the controls $\underline{\alpha}= (\alpha_1,\alpha_2,\alpha_3)^\intercal$ consist of two trap parameters: the stiffness ($\alpha_1>0$) and its location ($\alpha_2$), and the persistence of the OU process (with an effective equilibrium model being recovered for $\alpha_3 \rightarrow 0$). A strict passive regime is recovered for $v=0$.

\section{Results}
\label{sec:5}

\begin{figure}[t!]
    \centering
    \includegraphics[width=0.8\linewidth]{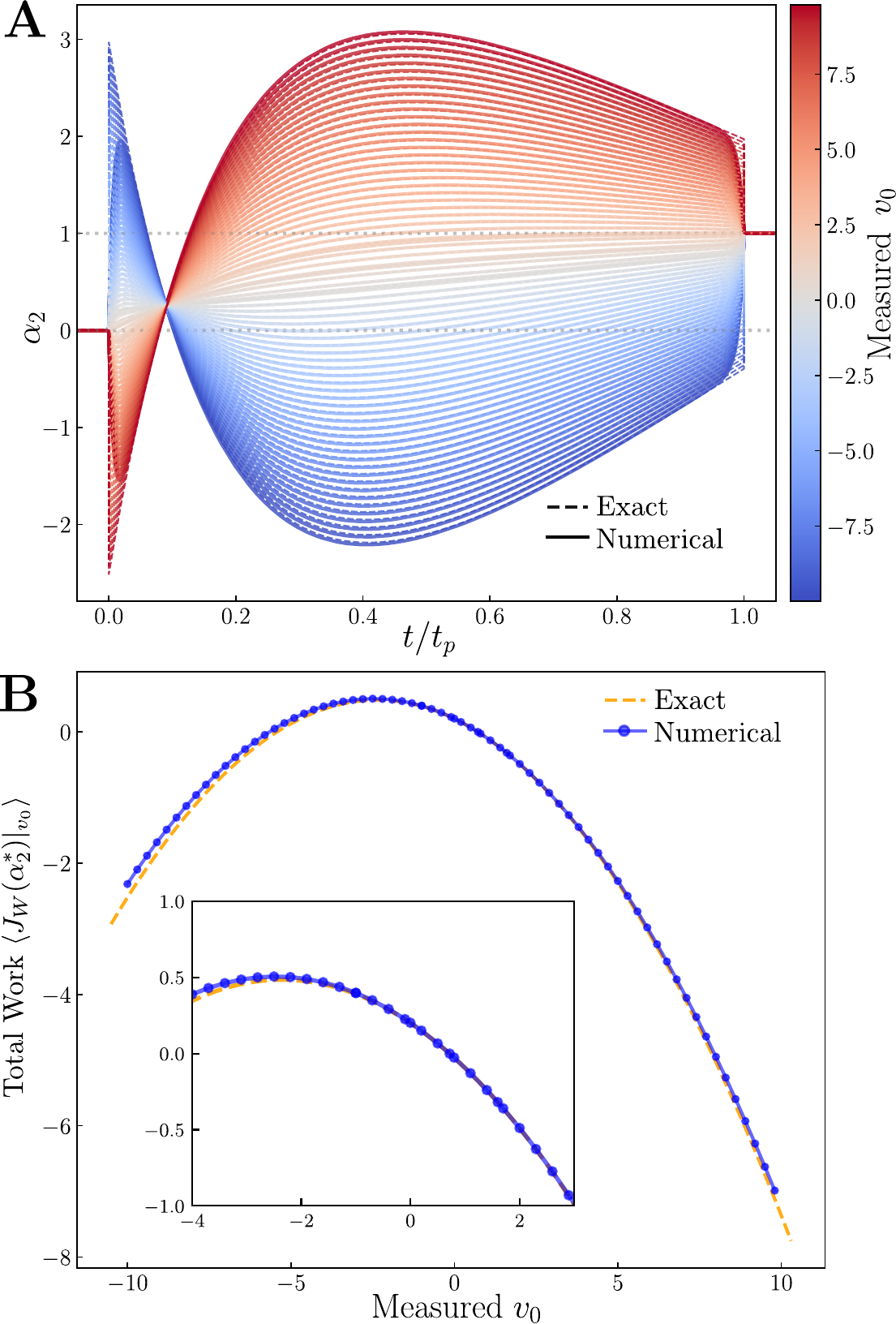}
    \caption{ \textbf{Comparing our method to the exact results of Sch\"{u}ttler et. al. (2025) \cite{Schuttler2025}.} \textbf{A} Work minimizing--closed-loop--trap center protocols ($\alpha_2$) as a function of the normalized protocol time, for different initial measurements of the self-propulsion $v_0$. The fanning of the protocols resembles a ``pirhana''. Numerical curves are for the regularized work ($J_R >0$). \textbf{B} Values of the closed-loop total work as a function of $v_0$ arising for both the exact and numerical optimal protocols. Model parameters: $\alpha_2(t=0^-) = 0$, $\alpha_2(t=t_p^+) = 1$, $\alpha_2(t) \in \mathbb{R}$, $t_p = 3$, $\alpha_1=1$, $D'=1$, and $\alpha_3 = 0.525$. Optimization parameters: $M=500$ and $m_\varepsilon = 10^{-4}$. }
    \label{fig:schuttlertest}
\end{figure}
\subsection{Single control}

First, we see how our numerical method to arrive at optimal protocols compares to single control problems that have been solved exactly. 

\textit{Equilibrium (Schmiedl-Seifert) test:-}A classic problem is the minimal-work open-loop (OL) control of a passive particle in a trap of varying stiffness \cite{Schmiedl2007}. In our model, this means fixing $\alpha_2 = 0$ (the origin for simplicity) and nullifying the self-propulsion $v = 0$ (making $\alpha_3$ irrelevant). For this problem we explore both the un-regularized thermodynamic cost ($J_R = 0$) and the regularized one, with moderately time-discretized controls. The precise cost function for this problem is:
\begin{align}
\langle {J}_W \rangle &= \frac{1}{2}\int_0^{t_p} dt \left(\dot{\alpha}_1(t) m_{r^2}(t)\right)\\
\langle \tilde{J}_W \rangle &= \frac{1}{2}\int_0^{t_p} dt \left(\dot{\alpha}_1(t) m_{r^2}(t) + m_\varepsilon \dot{\alpha}_1^2 \right),
\end{align}
where $m_\varepsilon = 10^{-4}$ in the second line, and the variance $m_{r^2}(t) = \langle r^2(t) \rangle$ obeys the closed ODE \cite{Davis2024}:
\begin{equation}
    \dot{m}_{r^2}(t) = -2 \mu \alpha_1(t) m_{r^2}(t) + 2D.
\end{equation}

Our optimal control protocols, for both the regularized and un-regularized work, converge on excellent agreement with the exact solutions for $t_p \gtrapprox 1$ (see Fig.~\ref{fig:schmeidltest}A). As expected, the regularized ($m_\varepsilon, J_R \neq 0$) controls deviate from the exact bang-bang protocols close to the end-points $t=0$ and $t=t_p$, with this deviation being rather large for small $t_p$ (\emph{e.g.,} $t_p = 0.1$). These smoothed protocols resemble those found in \cite{Zhong2024}, which also adopted a regularization scheme. For the chosen discretization of the protocols ($M = 500$), we observe significant chattering in the un-regularized, $J_R = 0$, controls close to the end-points (Fuller's phenomenon). Despite the chattering, and unlike the regularized protocols, the optimal un-regularized protocols appear to trace out the exact solution. Similar chattering behaviour was reported in \cite{Kamizaki2022}. Importantly, we find that the values of the work, taking all optimal protocols and injecting them into the raw--unregularized--work, are only a small amount higher (differences of $\sim 10^{-3}$) than the exact solution (see Fig.~\ref{fig:schmeidltest}B). This is somewhat surprising given the obvious differences in the fast numerical optimal controls compared with the exact analytical controls. We find that the regularized protocols result in the highest (least optimal) work values, which is explained by the presence of the regularization term in the cost \eqref{eq:RegGTRExact} and the fact that unregularized controls can leverage cost-free strong oscillations to reduce the work.

\textit{Active test:-} We next test our method for an active optimal closed-loop (CL) control problem: moving a trap from one location to another, given some initial measurement of the self-propulsion $v(t=0^-)=v_0$, at minimal work \cite{GarciaMillan2025,Schuttler2025}. We therefore work in the conditional ensemble where averaged quantities are conditioned on the initial (error-free) measurement. Given this, the deterministic mean for the self-propulsion reads as:
\begin{equation}
    m_v \vert_{v_0}(t) \equiv \langle v(t) \vert_{v_0} \rangle = v_0 \exp(-t/\alpha_3),
\end{equation}
and the ODE for the conditional mean position $m_r \vert_{v_0}(t) \equiv \langle r(t)\vert_{v_0}\rangle $ is
\begin{equation}
    \dot{m_r}\vert_{v_0}(t) = -\mu \alpha_1(m_r\vert_{v_0}(t) - \alpha_2(t)) + m_v\vert_{v_0}(t),
\end{equation}
where $\alpha_{1,3}$ are constant. The conditioned work is then written as:
\begin{equation}
    \langle J_W \vert_{v_0} \rangle = \alpha_1\int^{t_p}_0 dt~\dot{\alpha_2}(t)\left(\alpha_2(t) - m_r\vert_{v_0}(t) \right).
\end{equation}

Our numerical (regularized) protocols are in very good agreement with the exact optimal protocols (from \cite{Schuttler2025}), over a wide-range of initial measurements of the self-propulsion, with the exception of the discontinuous jumps at the end-points (see Fig.~\ref{fig:schuttlertest}A). While still being smooth, it is clear that the regularized protocols emulate the jumps at the end-points with protocol values at the first turning points being smaller (by a factor of $\approx 2/3$). Overlaying the protocols for different $v_0$s reveals an interesting fanning-out and re-intersection behaviour of the protocols (``pirhana''-shaped). This behaviour involves the following: (i) at earlier times, it is optimal to rapidly move the trap center towards (away from) +$\infty$ for $v_0 < 0$ ($v_0 > 0$); (ii) There is a re-intersection point of all the protocols at some $0 < t < t_p$; (iii) After the re-intersection time, the protocols then fan out again but now with the opposite behaviour to (i), \emph{i.e.,} it is now optimal to move the trap center away from (towards) +$\infty$  for $v_0<0$ ($v_0>0$). Importantly, these regularized (smoothed) protocols are almost as efficient as the bang-bang protocols (see Fig.~\ref{fig:schuttlertest}B), particularly for $|v_0| \leq 5$, with growing discrepancy for higher initial propulsion and more so for self-propulsion directed away from the trap target location.

\begin{figure}[t!]
    \centering
    \includegraphics[width=0.99\linewidth]{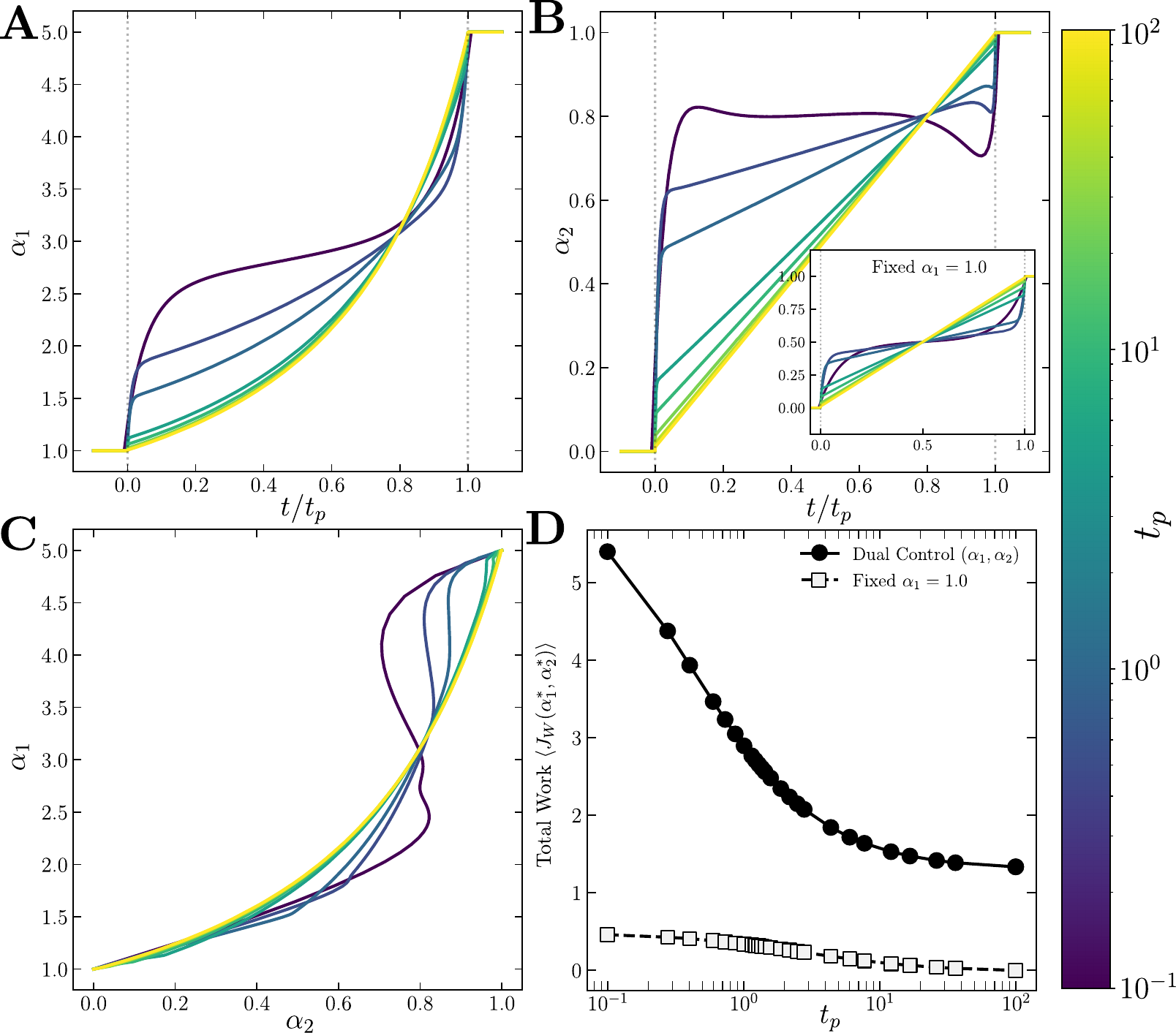}
    \caption{\textbf{Dual open-loop control of the trap stiffness and center at minimal work.} \textbf{A} Optimal stiffness protocols for various protocol durations, where the the stiffness is bounded by its end-point values. \textbf{B} Optimal center protocols with the protocols at fixed stiffness shown as an inset. \textbf{C} Control phase portrait for the protocols shown in (A,B). \textbf{D} Averaged total work for the optimal protocols in (A,B) compared to the open-loop single parameter (fixed $\alpha_1$) case. Model parameters: $\alpha_1(t=0^-) = 1$, $\alpha_1(t=t_p^+) = 5$, $\alpha_2(t=0^-) = 0$, $\alpha_2(t=t_p^+) = 1$, $\alpha_2(t) \in (0,1]$, $\alpha_1(t) \in (0,5]$, $t_p = 3$, $D'=2$, and $D=\mu=\alpha_3=1$. Optimization parameters: $M=1000$ and $m_\varepsilon = 10^{-4}$. }
    \label{fig:dc_ol_stiffnesslocation_work}
\end{figure}

\textit{Comparing to neural networks:-} Encouraged by our agreements with analytically exact results, we next sought to compare our gradient descent (AD) method for finding optimal thermodynamic controls with optimization via neural evolution \cite{Casert2024}. For this we focus on an active optimal control problem which--to the best of our knowledge--has no exact closed form solution: finding the optimal open-loop protocol for the trap stiffness at minimal dissipation ($J_H$) \cite{Davis2024,Casert2024}. 

For this particular problem the cost Lagrangian is given as
\begin{equation}
\begin{aligned}
        \mathcal{L}_H &= \langle \dot{\alpha}_1 \partial_{\alpha_1} \phi -\alpha_1 rv\rangle, \\
        &= \frac{\dot{\alpha}_1(t)}{2} m_{r^2}(t) - \alpha_1(t) m_{rv}(t),
\end{aligned}
\end{equation}
where $\alpha_2=0$ for simplicity and with the ODEs for the evolution of the moments going as:
\begin{equation}
    \begin{aligned}
        \dot{m}_{r^2}(t) &= m_{rv}(t) - \mu \alpha_1(t) m_{r^2}(t) + D, \\
        \alpha_3 \dot{m}_{rv}(t) &= D' - m_{rv}(t)(1 + \alpha_1(t) \mu \alpha_3 ).
    \end{aligned}
\end{equation}
The form of the boundary term in the cost function, $B_H$, depends whether or not one enforces a state-to-state (STS) transformation at $t=t_p^+$. Both cases are given below as:
\begin{align}
\label{eq:BH1}
B_H &= \frac{D' t_p}{\mu \alpha_3} + \frac{1}{2}\left((\alpha_1 m_{r^2})_{t=0} - (\alpha_1 m_{r^2})_{t=tp}\right), \nonumber \\
&= \frac{D' t_p}{\mu \alpha_3} + \frac{1}{2}\left(\frac{D'}{\mu(1+\alpha_1(0)\mu \alpha_3)} - (\alpha_1 m_{r^2})_{t=tp}\right), \\
B_H &\underset{\text{sts}}{=} \frac{D' t_p}{\mu \alpha_3} + \frac{1}{2}\left(\frac{D'}{\mu(1+\alpha_1(0)\mu \alpha_3)} - \frac{D'}{\mu(1+\alpha_1(t_p)\mu \alpha_3)} \right).
\end{align}

\begin{figure*}[t!]
    \centering
    \includegraphics[width=0.95\linewidth]{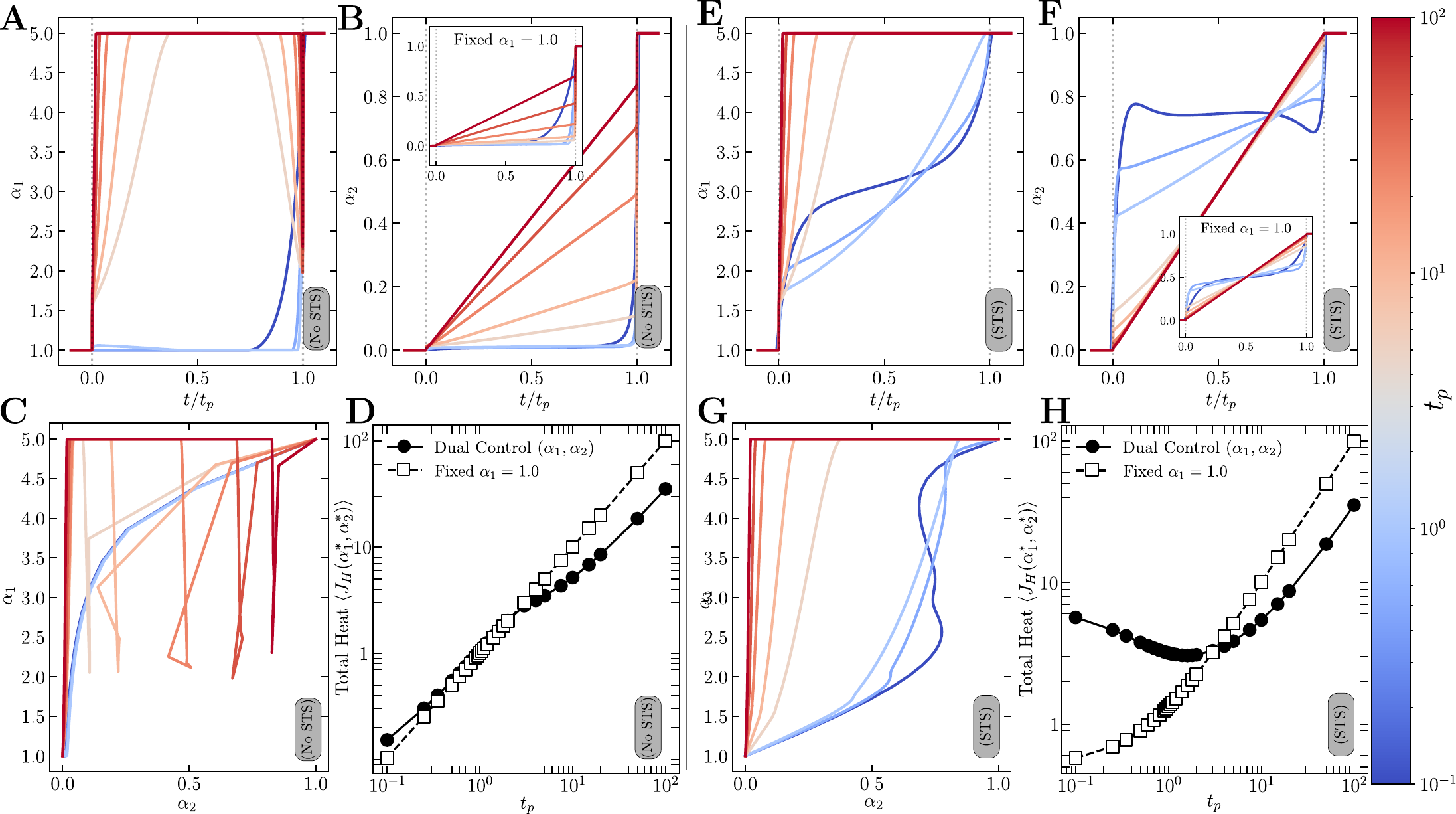}
    \caption{\textbf{Dual open-loop control of the trap stiffness and center at minimal heat dissipation.} \textbf{A} Optimal protocols for $\alpha_1$ as a function of the normalized protocol time $t/t_p$, for the case of no enforcement of a state-to-state transformation (No STS). \textbf{B} Optimal protocols for $\alpha_2$. Inset shows optimal $\alpha_2$ protocols at fixed trap stiffness. \textbf{C} Control phase portrait for the protocols shown in (A,B). \textbf{D} Averaged total heat for the optimal protocols, showing both the dual control and the single control costs. Open-loop dual control becomes more efficient at slower driving. \textbf{E-H} Same as (A-D) but for the case of enforcing an STS transformation. Model parameters: $\alpha_1(t=0^-) = 1$, $\alpha_1(t=t_p^+) = 5$, $\alpha_2(t=0^-) = 0$, $\alpha_2(t=t_p^+) = 1$, $\alpha_2(t) \in (0,1]$, $\alpha_1(t) \in (0,5]$, $t_p = 3$, $D'=2$, and $D=\mu=\alpha_3=1$. Optimization parameters: $M=1000$ and $m_\varepsilon = 10^{-4}$. }
    \label{fig:dc_ol_stiffnesslocation_heat}
\end{figure*}

For the same model parameters, constraints ($\alpha_1(0) \leq \alpha_1 (t) \leq\alpha_1(t_p)$), and control end-points our optimal protocols and costs are--on the whole--comparable with those derived from neural evolution \cite{Casert2024} (see Fig.~\ref{fig:casert}A-B), with some notable differences which we now discuss. Concerning the case of no STS transformation, for protocol durations $t_p < 5.74$ our protocols capture the same qualitative features as arrived at by neural evolution, such as non-monotonicity in some of the protocols, though with our protocol values being typically higher (at most $\approx25\%$, see Fig.~\ref{fig:casert}A). For the STS transformation protocols, we observe almost negligible difference from the neural evolution results (Fig.~\ref{fig:casert}). A major difference from \cite{Casert2024} is that, for the no STS transformation case, we find noticeable non-monotonicity in the protocols for all protocol durations larger than $t_p \gtrapprox 2$, whereas \cite{Casert2024} predicts monotonically increasing protocols for $t_p = 5.74$ and $t_p = 100$ (Fig.~\ref{fig:casert}A, C). We reveal that the time-extent of the non-monotonic feature, \emph{i.e.,} a rapid drop in trap stiffness followed by a more rapid climb up to the protocol end-point, decreases with increasing $t_p$ and is pushed towards $t \rightarrow t_p$ (Fig.~\ref{fig:casert}C). While the no STS protocols start to resemble the STS protocols for high $t_p$, we show that this late stiffness expansion and contraction remains the key difference. Only in the quasistatic limit, $t_p \rightarrow \infty$, would one expect the no STS protocols to converge onto the STS protocols. Despite the observed protocol differences, we show that the total costs (dissipation) are quantitatively similar (within $\approx15\%$) to \cite{Casert2024} (Fig.  \ref{fig:casert}D).

\subsection{Dual control}

We now explore active optimal control problems with the introduction of one more control degree of freedom (see Fig.~\ref{fig:Sketch}(ii)). We first look at open-loop control strategies that minimize the  dissipation or work, and then move onto closed-loop control strategies. From this point onward we only consider the regularized thermodynamic costs \eqref{eq:JRCost}.

\textbf{Open-loop}. \textit{Controlling trap stiffness and center:-} A natural extension of the single control problem of a trapped active particle is to have both the trap stiffness ($\alpha_1$) and center ($\alpha_2$) controllable (see Figs.~\ref{fig:Sketch}(ii), \ref{fig:dc_ol_stiffnesslocation_work} and \ref{fig:dc_ol_stiffnesslocation_heat}). For this dual control problem the objects defining the thermodynamic costs, for the work and the dissipation, are written as:
 \begin{align}
  \label{eq:OL_HW}
    \langle \mathcal{L}_W \rangle &= \frac{\dot{\alpha}_1(t)}{2}(m_{r^2}(t) - 2 \alpha_2(t) m_r(t) + \alpha_2^2(t)) \\
    & \quad - \alpha_1 (t) \dot{\alpha}_2 (t) (m_r(t) - \alpha_2(t)), \nonumber \\
     \langle \mathcal{L}_H \rangle &=  \langle \mathcal{L}_W(t) \rangle - \alpha_1(t) m_{rv}(t), \\
     B_H &= \frac{D'}{\mu \alpha_3} t_p   \label{eq:OL_HW2}\\ 
     &\quad - \left[ \frac{\alpha_1(t)}{2} \left( m_{r^2}(t) - 2 \alpha_2(t) m_r(t) + \alpha_2^2(t) \right) \right]_0^{t_p}, \nonumber
\end{align}
with the STS boundary term being the same as \eqref{eq:BH1}.

First, we focus on minimizing the (regularized) work. Unsurprisingly, we find that the stiffness protocol in the dual control case, where the trap center is also manipulated, is the same as in the single control case, where the trap center is fixed, given the translational invariance of the stiffness contribution to the work (Fig.~\ref{fig:dc_ol_stiffnesslocation_work}A). However, we find different optimal protocols for the trap-center control parameter, for simultaneous dual control, as compared to the single control protocols (Figs.~\ref{fig:dc_ol_stiffnesslocation_work}B,C). As the prescribed protocol for the stiffness is to tighten the trap ($\alpha_1: 1 \rightarrow 5$) we find that this increasing stiffness encourages (or permits) faster displacements for the trap center protocol. For shorter $t_p$, the trap center protocols become non-monotonic (Fig.~\ref{fig:dc_ol_stiffnesslocation_work}B)--a feature that has also been observed in other works \cite{Zhong2022,Whitelam2023, Zhong2024}. Additionally, the dual control trap-center protocol now loses its symmetry (along the $t/t_p = 1/2$ axis \cite{Loos2024}) compared with the fixed $\alpha_1$ single control, with the point of intersection (or ``meeting point'') of the curves being shifted towards the end of the protocol. For the explored protocol durations, we find that the total work expended is greater for the dual stiffness-center control as compared with the single control (fixed $\alpha_1$) case (Fig.~\ref{fig:dc_ol_stiffnesslocation_work}D). This finding is not so surprising as, given the prescribed end-points of the two control parameters, it is difficult to find protocols that compensate for the extra work in controlling an extra degree of freedom \eqref{eq:OL_HW}.

For this control problem we next considered minimizing the (regularized) total heat with no enforcement of a steady state at the end of the protocol (\emph{i.e.,} no STS transformation). As with the work, the trap stiffness protocols in the dual control case remain the same as in the fixed trap center case (see Figs.~\ref{fig:dc_ol_stiffnesslocation_heat}A and \ref{fig:casert}C). We find numerically different trap center protocols, compared with the single control at fixed $\alpha_1$, though, consistent with the minimal work problem, the overall form of the protocols remain largely the same (Fig.~\ref{fig:dc_ol_stiffnesslocation_heat}B). For the control parameter portrait, $\alpha_1(\alpha_2)$, we observe the appearance of loops for $t_p \gtrapprox 1$ (Fig.~\ref{fig:dc_ol_stiffnesslocation_heat}C). We note that such loops in the parameter portraits have been observed in another work that has explored similar dual control problems \cite{Casert2024}. Interestingly, for fast protocols ($t_p \lessapprox 1$) we find that open-loop dual control emits slightly more heat as compared to the fixed $\alpha_1$ single control, though for slower protocols ($t_p \gtrapprox 1$) the dual control becomes more efficient (Fig.~\ref{fig:dc_ol_stiffnesslocation_heat}D). We believe this extra efficiency at higher $t_p$ has its origin in the late expansion (breathing) of the trap stiffness which emerges only for slow protocols (see Figs.~\ref{fig:dc_ol_stiffnesslocation_heat}A,C).

Imposing a state-to-state (STS) transformation results in markedly different protocols for both the trap stiffness and trap center (see Figs.~\ref{fig:dc_ol_stiffnesslocation_heat}E-G), which is consistent with the single-control (fixed $\alpha_2$) case (Fig.~\ref{fig:casert}C). We find that the heat-optimal $\alpha_1$ protocols match the work-optimal $\alpha_1$ protocols for short $t_p$ (Fig.~\ref{fig:dc_ol_stiffnesslocation_heat}E and \ref{fig:dc_ol_stiffnesslocation_work}A respectively). This can be explained upon inspection of the form of the costs (\ref{eq:OL_HW}-\ref{eq:OL_HW2}) where imposing small $t_p \ll 1$ results in $B_H \approx 0$ and $\langle \mathcal{L}_H \rangle \approx \langle \mathcal{L}_W \rangle$, and so the thermodynamic costs become almost the same. For the trap center, the heat-optimal and work-optimal protocols appear almost the same for the range of $t_p$ explored here (Fig.~\ref{fig:dc_ol_stiffnesslocation_heat}F and \ref{fig:dc_ol_stiffnesslocation_work}B respectively). This, again, can be explained by inspection of the cost functions, and is mainly due to the difference in heat and work Lagrangians, $-\alpha_1(t) m_{rv} (t)$, having no $\alpha_2(t)$ dependence. For the STS case, the control phase portrait does not exhibit loops, in contrast to the no STS transformation case, largely due to the monotonic behaviour of $\alpha_1(t)$ (Fig.~\ref{fig:dc_ol_stiffnesslocation_heat}G). The total heat resulting from dual control is, again, costlier for shorter protocols but now by at most an order of magnitude (for $t_p \approx 0.1$) (Fig.~\ref{fig:dc_ol_stiffnesslocation_heat}H). As with the no STS case, there is a crossover point for $t_p \approx 1$ where the total heat dissipated becomes markedly less for dual parameter control with efficiency gains similar to the no STS case. The main lesson here is that for open-loop dual control of the stiffness and trap center, less (compared with single control) heat is expelled for slower protocols compared to shorter ones.
\begin{figure}[t!]
    \centering
    \includegraphics[width=0.99\linewidth]{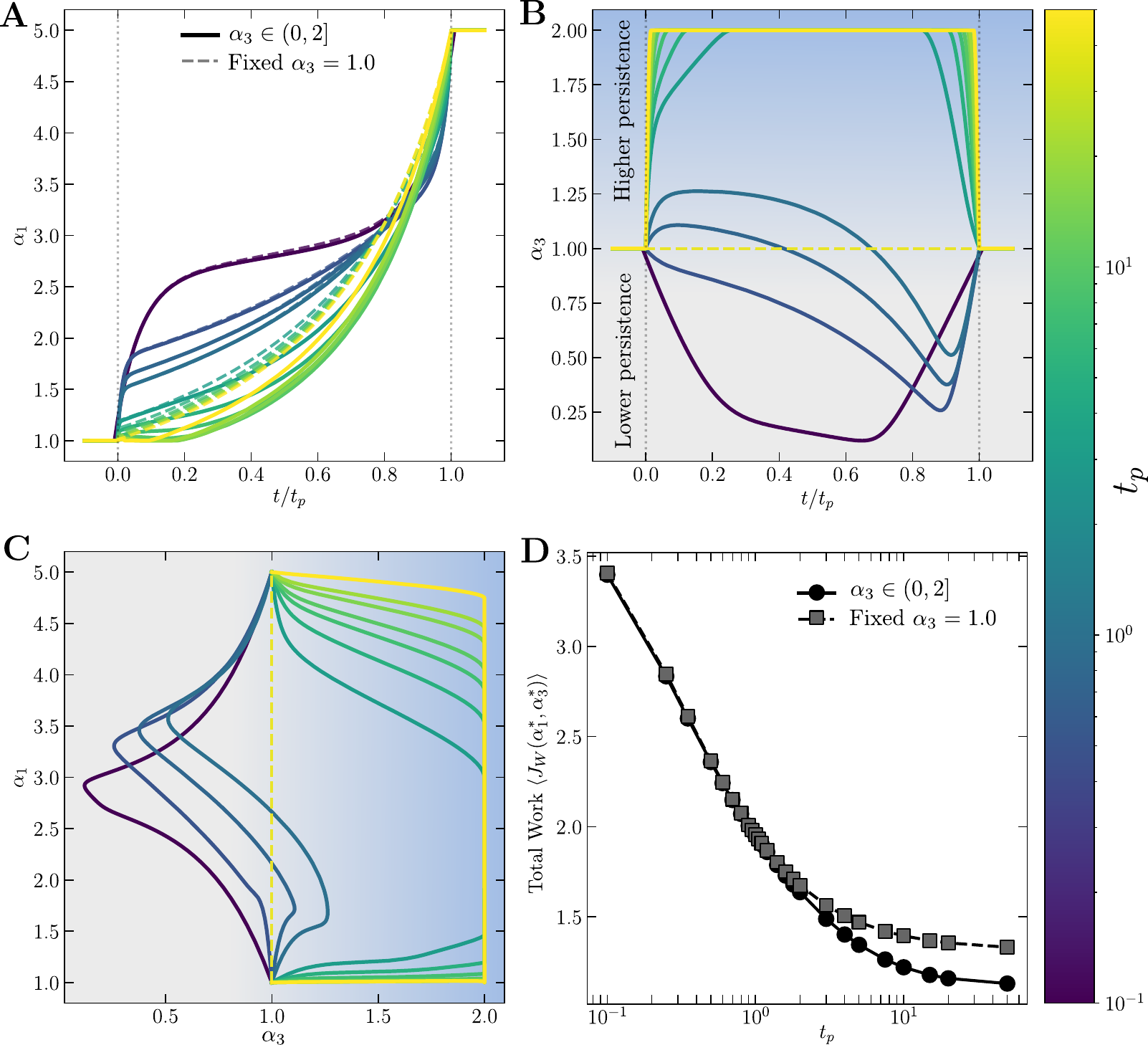}
    \caption{ \textbf{Dual open-loop control of the trap stiffness and the (``flexible'') persistence time at minimal work.} \textbf{A} Optimal protocols for the trap stiffness $\alpha_1$ against $t/t_p$ for various protocol durations $t_p$. The solid line represents the optimal stiffness protocol with a flexible persistence $\alpha_3$, and the dotted line is for fixed $\alpha_3$. \textbf{B} Optimal persistence $\alpha_3$ protocols as a function of $t/t_p$. Values above $\alpha=1$ are considered more active, and values below this value are less active (more passive). Flexibility here means that the end-points are fixed at the same value with some freedom ($\pm 1$) to go above and below the end-point value. \textbf{C} Control phase portrait for the controls shown in (A,B). \textbf{D} Total dual control OL work for the flexible and fixed $\alpha_3$ cases. Model parameters: $\alpha_1(t=0^-) = 1$, $\alpha_1(t=t_p^+) = 5$, $\alpha_3(t=0^-) = \alpha_3(t=t_p^+) = 1$, $\alpha_1(t) \in (0,5]$, $\alpha_3(t) \in (0,2]$, $D'=2$, and $D=\mu=1$. Optimization parameters: $M=1000$ and $m_\varepsilon = 10^{-4}$. }
    \label{fig:dc_ol_stiffnesspersistence_work}
\end{figure}

\begin{figure*}[t!]
    \centering
    \includegraphics[width=0.99\linewidth]{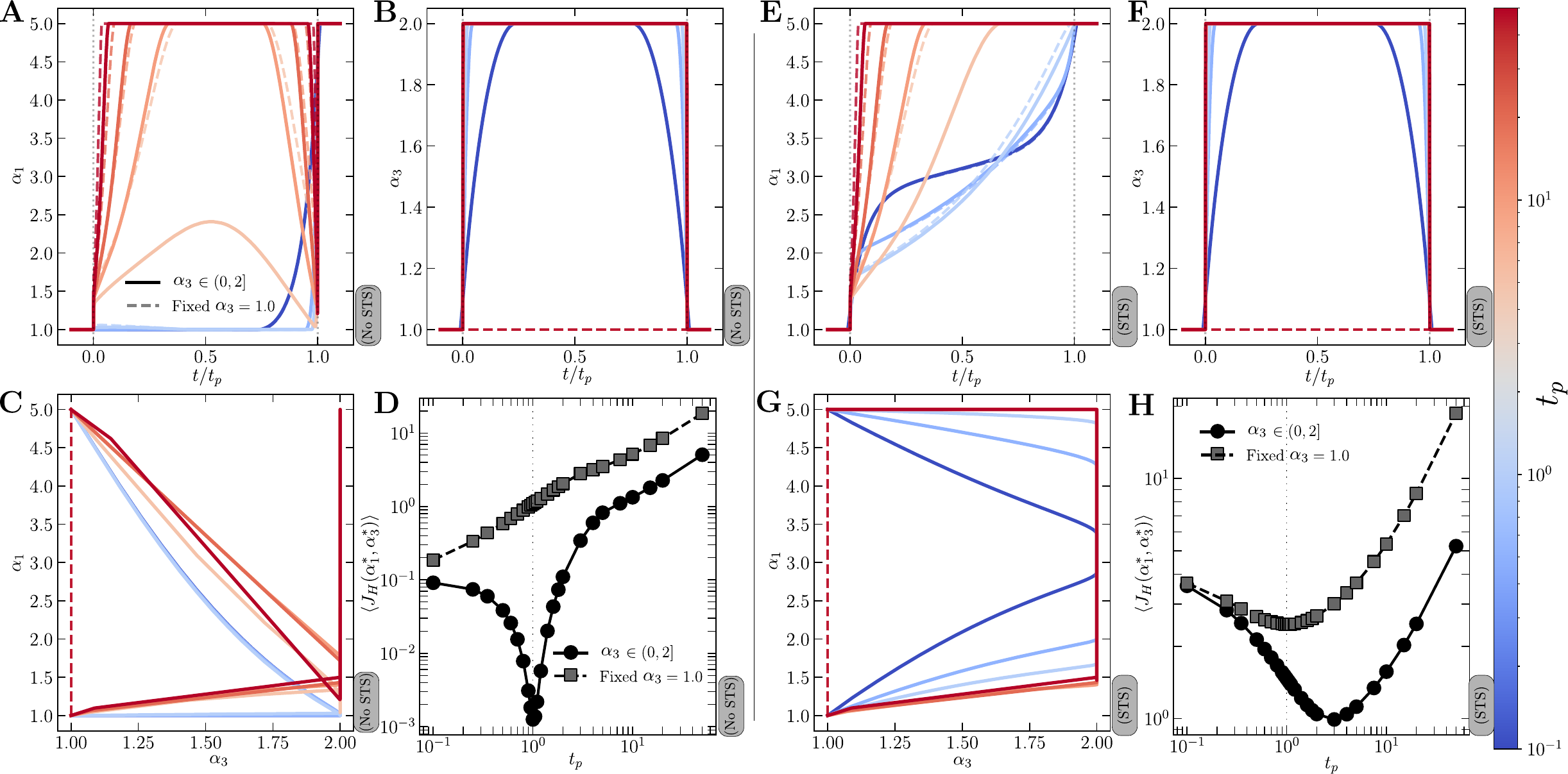}
    \caption{ \textbf{Dual open-loop control of the trap stiffness and persistence time at minimal dissipation.}  \textbf{A} Optimal $\alpha_1$ as a function of $t/t_p$, shown for flexible (solid lines) and fixed (dashed lines) persistence. \textbf{B} Optimal $\alpha_3$ as a function of $t/t_p$, note the symmetry along the $t/t_p = 1/2$ axis and the preference for higher persistence. \textbf{C} Control phase portrait. \textbf{D} Total heat as a function of $t_p$ for the flexible and fixed $\alpha_3$ cases. The vertical dotted line marks the $t_p = \alpha_3( t_p || 0)$. \textbf{E-H} The same as (A-B) but with a state-to-state (STS) transformation imposed.  Model parameters: $\alpha_1(t=0^-) = 1$, $\alpha_1(t=t_p^+) = 5$, $\alpha_3(t=0^-) = \alpha_3(t=t_p^+) = 1$, $\alpha_1(t) \in (0,5]$, $\alpha_3(t) \in (0,2]$, $D'=2$, and $D=\mu=1$. Optimization parameters: $M=1000$ and $m_\varepsilon = 10^{-4}$. }
    \label{fig:dc_ol_stiffnesspersistence_heat}
\end{figure*}

\textit{Controlling trap stiffness and persistence:-} We next wondered how having control of the activity affects the efficiency of controlling the trap, compared to fixed activity. Although some works have explored controlling the activity of a trapped particle, this natural question has yet to be answered. To this end, we explore a dual control problem where the stiffness $\alpha_1$ is driven and the persistence time $\alpha_3$ of the active OU process is made ``flexible'' between two end-points which are fixed and made the same, \emph{i.e.,} $\alpha_3(0) = \alpha_3(t_p) = 1$ and $\alpha_3 (t) \in (0,2] \equiv \alpha_3(0) \pm 1 $. Indeed, given the end-points are the same one can think of this as a cycle in the activity. The mathematical objects entering the thermodynamic cost functions for this control problem are given as:
\begin{align}
    \langle \mathcal{L}_W \rangle &= \frac{\dot{\alpha}_1(t)}{2} m_{r^2}(t), \label{eq:SP_LW} \\
    \langle \mathcal{L}_H \rangle &= \langle \mathcal{L}_W \rangle - \alpha_1(t) m_{rv}(t) + \frac{D'}{\mu \alpha_3(t)},\label{eq:SP_LH}  \\
    B_H &= - \left[ \frac{\alpha_1(t)}{2} m_{r^2}(t) \right]_0^{t_p}, \\
    B_H &\underset{sts}{=} \frac{D'}{2\mu} \left( \frac{1}{1 + \mu \alpha_1(0) \alpha_3(0)} - \frac{1}{1 + \mu \alpha_1(t_p) \alpha_3(t_p)} \right) \label{eq:SP_BH_STS},
\end{align}
where, given the now non-constant $\alpha_3(t)$, some of the terms that were previously in $B_H$ are now part of the running-cost.

For open-loop control with the work as the cost, we find that having a protocol on $\alpha_3$ changes (as compared to fixed $\alpha_3$) the numerical optimal protocols for $\alpha_1$, with this deviation growing for increasing $t_p$ (see Fig.~\ref{fig:dc_ol_stiffnesspersistence_work}A). Interestingly, we find two contrasting regimes for the optimal protocols for the persistence time $\alpha_3$: (i) for $t_p < 1$ the optimization prefers to reduce the persistence (hence degree of activity) for the majority of the protocol duration, and (ii) for longer $t_p \gtrapprox 1$ it is more optimal to rapidly increase the persistence to the upper bound for almost all of the protocol duration, and only reduce it to the end-point in the last $\sim 1\%$ of the protocol (see Fig.~\ref{fig:dc_ol_stiffnesspersistence_work}B). In other words, for faster protocols it is more optimal to reduce activity (move towards passiveness) in contrast to slower protocols where it pays to be as persistent as possible. We note that at intermediary protocol durations we find that the persistence is first decreased and then increased towards the end of the protocol. The control phase portrait shown in Fig.~\ref{fig:dc_ol_stiffnesspersistence_work}C offers a clue of an explanation: as $t_p \rightarrow 0$ the optimizer drives the persistence to its lowest value at the approximate center-point of the trap stiffness $(\alpha_1(t_p) - \alpha_1(0))/2$. Importantly, the total work is lower for the dual control (flexible $\alpha_3$) case as compared to the fixed persistence case (Fig.~\ref{fig:dc_ol_stiffnesspersistence_work}D), though the gains in efficiency become negligible at short durations $t_p \lessapprox 1$. So for rapid protocols, it is probably more practical to perform the simplest protocol consisting of solely controlling $\alpha_1$.

\begin{figure*}[t!]
    \centering
    \includegraphics[width=0.85\linewidth]{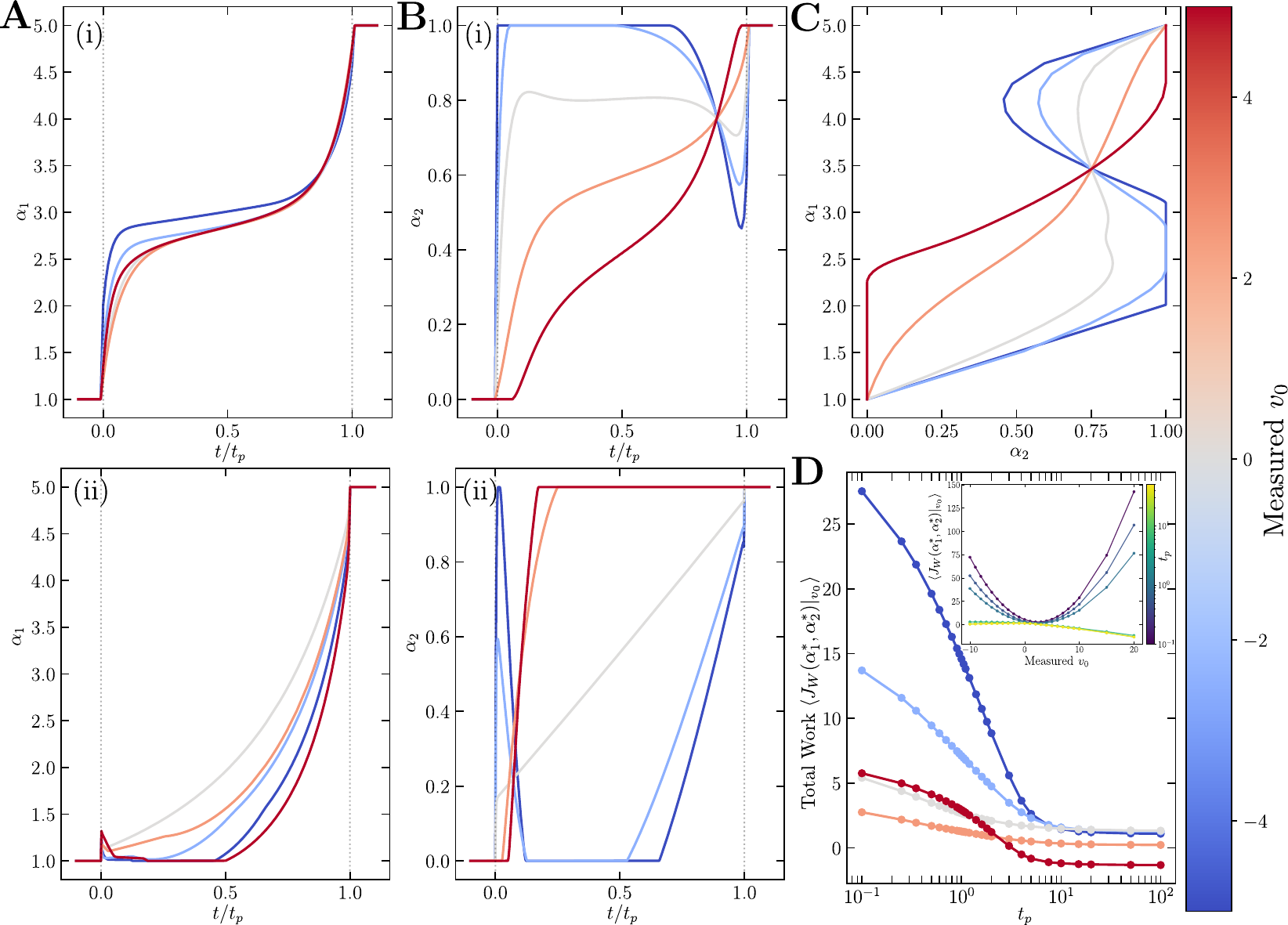}
    \caption{\textbf{Dual closed-loop control of the trap stiffness and center accounting for initial (perfect and zero-cost) measurement of the self-propulsion.} \textbf{A} Optimal stiffness $\alpha_1$ protocols as a function of $t/t_p$ at various $v_0$ measurements and shown for two protocol durations (i) $t_p = 0.1$, and (ii) $t_p = 5.0$. \textbf{B} Optimal trap center $\alpha_2$ protocols for various $v_0$, again, we show protocols for (i) $t_p=0.1$ and (ii) $t_p = 5.0$. The ``pirhana'' behaviour, as was first shown in Fig.~\ref{fig:schuttlertest}A, is still observable here. \textbf{C} The control phase portrait, note the re-intersection point. \textbf{D} The CL total work as a function of $t_p$ for various $v_0$. The inset shows the work as a function of $v_0$ for selected $t_p$. Model parameters: $\alpha_1(t=0^-) = 1$, $\alpha_1(t=t_p^+) = 5$, $\alpha_2(t=0^-) = 0$, $\alpha_2(t=t_p^+) = 1$, $\alpha_1(t) \in (0,5]$, $\alpha_2(t) \in (0,1]$, $D'=2$, $v_0 \in (-5,5)$, and $D=\mu=\alpha_3=1$. Optimization parameters: $M=1000$ and $m_\varepsilon = 10^{-4}$. }
    \label{fig:dc_cl_stiffnesslocation_work}
\end{figure*}

In contrast to the work, for the total heat we find that making $\alpha_3$ flexible offers significant gains in efficiency (Fig.~\ref{fig:dc_ol_stiffnesspersistence_heat}). First, we observe almost no differences to the fixed $\alpha_3$ stiffness protocols, even at large $t_p$ (Fig.~\ref{fig:dc_ol_stiffnesspersistence_heat}A). However, we do note a growing--though still only small--deviation at larger $t_p$, with this being significantly smaller than was found for the work-optimal protocols (Fig.~\ref{fig:dc_ol_stiffnesspersistence_work}A). Interestingly, for the dissipation, it appears that the optimal strategy is to always ramp up the persistence and then drop it back down again, with longer protocols favoring more abrupt changes in $\alpha_3$ (Fig.~\ref{fig:dc_ol_stiffnesspersistence_heat}B,C). In contrast to the asymmetric (along the $t/t_p = 1/2$ axis) work-optimal persistence protocols, the heat-optimal persistence protocols exhibit a clear symmetry. This fundamental difference arises from the mathematical structure of the Lagrangians (\ref{eq:SP_LW},\ref{eq:SP_LH}). The work cost $\langle \mathcal{L}_W \rangle \propto \dot{\alpha}_1 m_{r^2}$ is driven by the mechanical compression of the trap ($\dot{\alpha}_1 > 0$), imposing an arrow of time. To prevent a late-stage spike in the variance $m_{r^2}$ from incurring a massive work penalty, the reduction of $\alpha_3$ to its final boundary condition must be maximally delayed, resulting in a skewed, asymmetric protocol. Conversely, the heat cost $\langle \mathcal{L}_H \rangle$ is dominated by the continuous housekeeping heat ${D'/(\mu \alpha_3(t))}$. Because this running cost is state-independent and carries no directional $\dot{\alpha}$ dependence, its minimization strictly dictates maximizing the area under $\alpha_3(t)$. The optimal path to satisfy the identical end-points $\alpha_3(0)=\alpha_3(t_p)$ while suppressing this dominant $1/\alpha_3$ penalty is naturally a symmetric pulse, masking the underlying asymmetry of the mechanical compression from a stiffer trap. 

Rather surprisingly, we find that heat-minimizing dual controls, with a flexible persistence parameter, is always more efficient than for single control at fixed persistence (Fig.~\ref{fig:dc_ol_stiffnesspersistence_heat}D). Crucially, we find the maximum efficiency gains, up to four orders of magnitude, when the protocol duration matches the end-point value of the persistence $t_p/\alpha_3 = 1$ with rather steep climbs up in dissipation in either direction $t_p < 1$ or $t_p > 1$. Figs.~\ref{fig:dc_ol_stiffnesspersistence_heat}E-G show the optimal stiffness and persistence protocols for the case of an enforced STS transformation. As with the no STS case, we find that the $\alpha_1(t)$ optimal protocols do not differ greatly from the optimal protocols at fixed persistence (Fig.~\ref{fig:dc_ol_stiffnesspersistence_heat}E). The optimal persistence protocols are found to be exact replicas of the no STS transformation protocols (Fig.~\ref{fig:dc_ol_stiffnesspersistence_heat}E), which is expected given that the cost terms containing $\alpha_3$ remain the same for both cases (\ref{eq:SP_LH}, \ref{eq:SP_BH_STS}). Gratifyingly, the positive efficiency gains in having an optimal control on the (flexible) persistence parameter is robust even with a STS transformation enforced, and that there is a protocol duration which confers a rather large efficiency gain (Fig.~\ref{fig:dc_ol_stiffnesspersistence_heat}H). Interestingly, the optimal protocol duration, defined as $\langle J_H(t_p^\ast) \rangle < \langle J_H(t_p^\ast) \rangle$ \cite{Davis2024}, for the dual control problem ($t_p^\ast \approx 3$) is shifted slightly ahead of the optimal protocol duration for the single control at fixed persistence ($t_p^\ast = \alpha_3 = 1$), which we largely attribute to the fact that this--now-controllable--active time-scale is $\approx 2$-fold larger for the majority of the protocol (Fig.~\ref{fig:dc_ol_stiffnesspersistence_heat}F).

\textbf{Closed-loop}. Now we turn to closed-loop (CL) dual control, where we make an initial measurement of the self-propulsion $v_0$ as was done for single parameter control (as shown in Fig.~\ref{fig:schuttlertest}). For this closed-loop control, we focus solely on minimizing the total work $\langle J_W \rangle$. As was done for the OL dual control, we impose lower and upper bounds on the control parameters. For the trap stiffness and center the lower are the start- and end-points respectively.

\textit{Controlling trap stiffness and center:-} Given that we are now working in the conditional ensemble, the work is slightly different and, for control on $\alpha_{1,2}$, is now written as:
\begin{align}
    \langle \mathcal{L}_W \vert_{v_0}\rangle  &= \frac{\dot{\alpha}_1(t)}{2}\left( m_{r^2}\vert_{v_0}(t) - 2 \alpha_2(t) m_r\vert_{v_0}(t) + \alpha_2^2(t) \right) \\
    & \quad - \alpha_1 (t) \dot{\alpha}_2 (t) \left( m_r\vert_{v_0}(t) - \alpha_2(t) \right). \nonumber
\end{align}

The optimal protocols for the dual and closed-loop control problem, for the explored range of $v0 = (-5, 5)$, that minimize the work are shown in Figs.~\ref{fig:dc_cl_stiffnesslocation_work}A-C. For a measurement of zero self-propulsion ($v_0 = 0$) we find that the optimal trap stiffness and center protocols replicate those found for dual open-loop control, shown in Fig.~\ref{fig:dc_ol_stiffnesslocation_work}, which is expected given that $\langle v \rangle=0$. Measurement of a non-zero self-propulsion $v_0 \neq 0$ leads to protocols that are numerically distinct from the OL case, with these protocols still retaining a form that is qualitatively similar to the zero measurement case. Specifically for the stiffness, initial knowledge of the activity results in a wider divergence of protocols for longer protocol durations (Figs.~\ref{fig:dc_cl_stiffnesslocation_work}A(i-ii)). However, for the same range of protocol durations, we see a greater ``fanning out'' of protocols for the trap center, $\alpha_2$, as compared with the trap stiffness, $\alpha_1$ (Fig.~\ref{fig:dc_cl_stiffnesslocation_work}B(i-ii), and see also the single control protocols Fig.~\ref{fig:schuttlertest}).  As with the single CL trap center protocols, we observe similar ``pirhana'' behaviour: a fanning out of protocols, with greater initial trap movement towards $+\infty$ for $v_0<0$, with the protocols re-intersecting at some $0 < t < t_p$ and then the protocols fan out once more with a late retraction of the trap for $v_0<0$. Despite this qualitative similarity to the single CL case, we note that a major difference in the shapes is caused by the constraint $\alpha_2(t) \in (0,1)$ which clips trap center movement towards $\pm \infty$ and results in monotonic (in $t/t_p$) protocols for $v_0> 0$.

Regarding the CL dual control work, adding a protocol for increasing the trap stiffness results in higher costs for larger $|v_0|$ for short $t_p \lessapprox 3$ (Fig.~\ref{fig:dc_cl_stiffnesslocation_work}D and inset). Increasing $t_p \gtrapprox 3$ changes this behaviour, with the least work occurring for large and positive $v_0$ but with greater work for large and negative $v_0$. In fact, for long enough durations ($t_p \gtrapprox 3$) the controller extracts work from the system, albeit at much smaller values than  was seen in the single (trap center) control case (see Fig.~\ref{fig:schuttlertest}B). We find that, as compared to the OL control, barely any efficiency gains can be obtained in the total control work for large and negative $v_0$ measurements. This is useful information, given that we do not account for the cost of the $v_0$ measurement which would make the CL $v_0 < 0$ work even higher \cite{Schuttler2025}. For long protocol durations $t_p \gtrsim 10 \alpha_3$ we find that CL control is--on average--more efficient than OL control; The mathematical reason behind this is the following: $\langle J_W \vert_{v_0 > 0} \rangle_{CL} < \langle J_W \rangle_{OL} \gtrapprox \langle J_W \vert_{v_0 < 0} \rangle_{CL}$ for long $t_p$ (see Fig.~\ref{fig:dc_cl_stiffnesslocation_work}D inset). For short $t_p$ the OL control wins out, as the previous relationship inverts, \emph{i.e.,} $\langle J_W \vert_{v_0 > 0} \rangle_{CL} \gtrapprox \langle J_W \rangle_{OL} < \langle J_W \vert_{v_0 < 0} \rangle_{CL}$.

\begin{figure}[t!]
    \centering
    \includegraphics[width=0.99\linewidth]{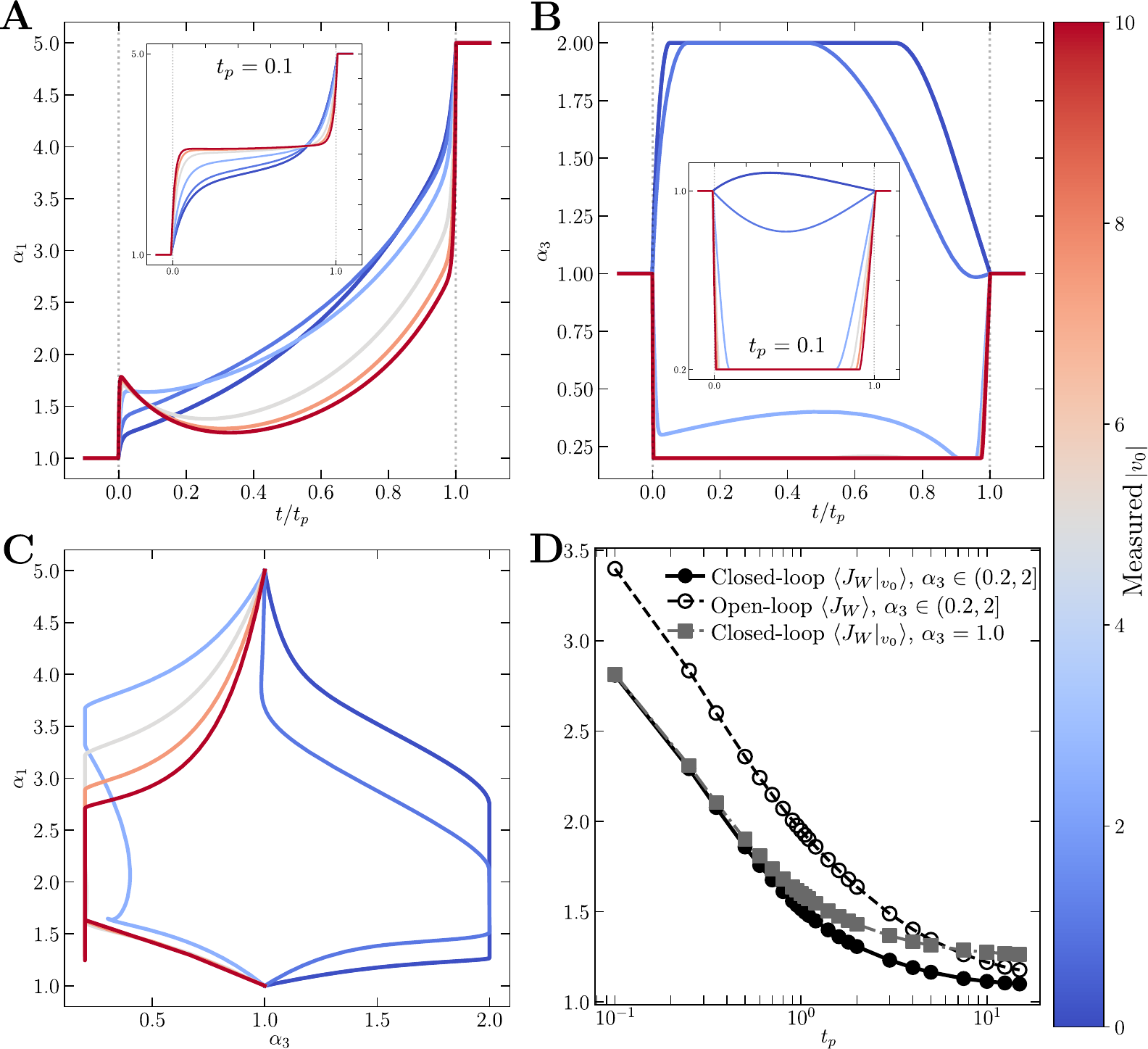}
    \caption{\textbf{Dual closed-loop control of the trap stiffness and persistence parameters at minimal work}. \textbf{A} Optimal trap stiffness against $t/t_p$ shown for different measured self-propulsion magnitudes $|v_0|$. We show the magnitude of $v_0$ due to the symmetry in the cost function (see main text). \textbf{B} Optimal flexing of the persistence parameter against $t/t_p$. \textbf{C} Control phase portrait. \textbf{D} Comparing the total work for carrying out the optimal protocols, as a function of $t_p$. We show: the CL (conditioned) work for flexible persistence (black solid line, filled circles), the CL work at fixed persistence (dark grey line, filled squares), and the work for the OL flexible persistence case. The CL flexible persistence controls are the most optimal across $t_p$. Model parameters: $\alpha_1(t=0^-) = 1$, $\alpha_1(t=t_p^+) = 5$, $\alpha_2(t=0^-) = 0$, $\alpha_2(t=t_p^+) = 1$, $\alpha_1(t) \in (0,5]$, $\alpha_2(t) \in (0,1]$, $D'=2$, $v_0 \in (-5,5)$, $t_p=1$ (insets $t_p = 0.1$), and $D=\mu=\alpha_3=1$. Optimization parameters: $M=1000$ and $m_\varepsilon = 10^{-4}$. }
    \label{fig:dc_cl_stiffnesspersistence_work}
\end{figure}

\textit{Controlling trap stiffness and persistence:-} We next wondered whether the efficiency gains from making the persistence parameter $\alpha_3$ flexible, that we observed for the OL dual control case (Fig.~\ref{fig:dc_ol_stiffnesspersistence_work}), carried over to CL dual control. This problem is of a wholly different character than has currently been explored, as it involves making measurements of the activity (self-propulsion) and then having the ability to change that activity going forward. The function that defines the work in this CL setup is:
\begin{align} \label{eq:dc_cl_SP_work}
    \langle \mathcal{L}_W \vert_{v_0} \rangle &= \frac{\dot{\alpha}_1(t)}{2} m_{r^2}\vert_{v_0}(t; \alpha_3(t)).
\end{align}
For this problem, we make one small adjustment to the control bounds: we increase the lower $\alpha_3$ bound from 0 to 0.2. We do this as we found that, in some cases, the optimizer tried to make the system passive through $\alpha_3 \rightarrow 0$, but this caused numerical problems for very small values of $\alpha_3$ which result in very large values of the cost \eqref{eq:dc_cl_SP_work}. Because the trap center remains fixed at the origin, the harmonic potential is perfectly symmetric. Consequently, the conditional spatial variance, which entirely dictates the mechanical work, depends solely on the squared initial velocity. Therefore, the thermodynamic cost is perfectly symmetric with respect to the initial direction of the self-propulsion, meaning only its magnitude determines the optimal protocols and the resulting work.

First, the optimal protocols shown in Fig.~\ref{fig:dc_cl_stiffnesspersistence_work}A-C are, unsurprisingly, different for various $|v_0|$. For short control durations, trap stiffness protocols for small $|v_0|$ converge to the open-loop protocols (Figs.~\ref{fig:dc_cl_stiffnesspersistence_work}A (inset) and \ref{fig:dc_ol_stiffnesspersistence_work}A). As for the stiffness-center CL dual control, we find non-monotonic $\alpha_1$ protocols for longer protocol durations. Interestingly, for low initial activity ($|v_0| \lesssim 2-3$) the optimal controls involve a rapid increase (relative to its default value) of the persistence parameter $\alpha_3$ which is maintained for the majority of the protocol duration (Fig.~\ref{fig:dc_cl_stiffnesspersistence_work}B). Contrastingly, high initial $|v_0|$ results in the opposite effect: a rapid decrease of $\alpha_3$ and then only in the very late stages of the protocol (say the last $\%5$) is the activity brought back to unity. This suggests that there is an optimal-level of activity for efficient control, which the controller can exploit via the combination of initial measurement and control of persistence. Indeed, overall it is beneficial to have both more information, via initial measurements of the activity, and flexibility, in the degree of activity, as shown through lower work values as compared to the CL fixed-$\alpha_3$ and OL flexible $\alpha_3$ cases (Fig.~\ref{fig:dc_cl_stiffnesspersistence_work}D). As with the OL case, for short $t_p$ the benefits of flexible persistence are marginal. We note that, at some protocol duration (here $t_p \approx 5-7$ for the chosen parameters) the OL flexible-$\alpha_3$ control becomes more efficient than the CL fixed-$\alpha_3$ control.

\begin{figure*}[t!]
    \centering
    \includegraphics[width=0.95\linewidth]{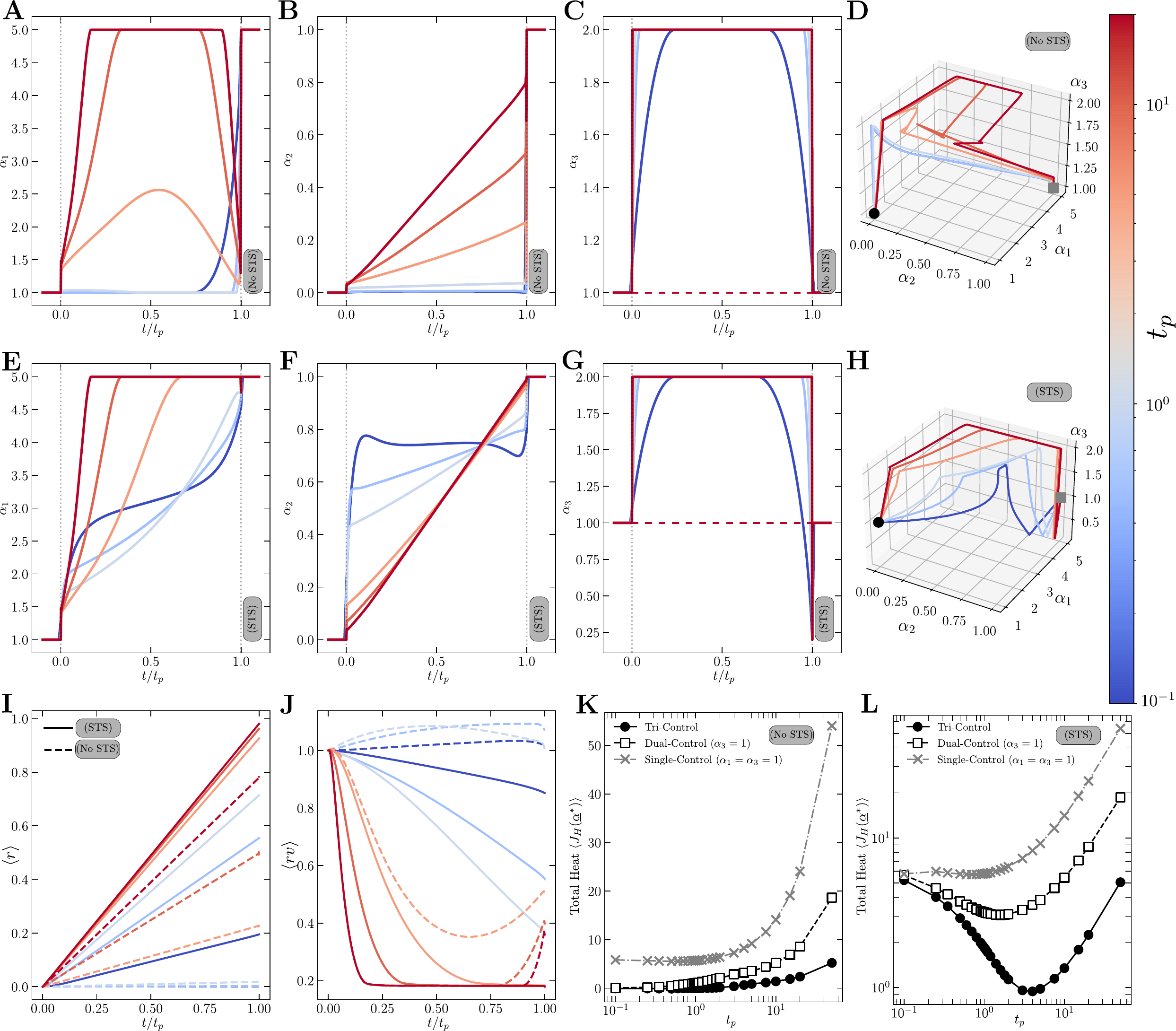}
    \caption{\textbf{Open loop tri-control at minimal dissipation}. \textbf{A} Optimal protocols for the trap stiffness as a function of $t/t_p$, for selected $t_p$s, for no STS transformaton. \textbf{B} Optimal trap translocation protocols. \textbf{C} Optimal persistence protocols, with the dashed line the fixed $\alpha_3=1$ case. \textbf{D} The tri-control phase portrait, showing where protocols begin (black filled circle) and end (dark grey filled square). \textbf{E-H} Same as (A-D) but for an imposed STS transformation. \textbf{I} The averaged particle position ($m_r \equiv \langle r \rangle$) as a function of $t/t_p$ for an STS transformation (solid lines) or no STS transformation (dashed lines). For fast protocols the particle lags behind. \textbf{J} Averaged product of the particle position times the self-propulsion, which appears in the costs as $m_{rv}$, as a function of $t/t_p$. Model parameters: $\alpha_1(t=0^-) = 1$, $\alpha_1(t=t_p^+) = 5$, $\alpha_2(t=0^-) = 0$, $\alpha_2(t=t_p^+) = 1$, $\alpha_3(t=0^-) = \alpha_3(t=t_p^+) = 1$, $\alpha_1(t) \in (0,5]$, $\alpha_2(t) \in (0,1]$, $\alpha_3(t) \in (0.2,2]$, $D'=2$, and $D=\mu=1$. Optimization parameters: $M=1000$ and $m_\varepsilon = 10^{-4}$. }
    \label{fig:tc_heat}
\end{figure*}

\subsection{Tri-control}

Lastly, we explore the most involved optimal thermodynamic control problems considered here: the simultaneous control of three model parameters (see Fig.~\ref{fig:Sketch}(iii)). It is worth reminding ourselves of the physical scenario being considered, and which motivates this section. We are considering the translocation of some active particle, whose degree of activity can be measured and varied, using a particle trap whose tightness and location can be precisely manipulated. As with the other sections we focus on performing our control as efficiently as possible, with the meaning of efficiency given to us by the adopted definitions of work or heat dissipation (\ref{eq:SekimotoHeat}-\ref{eq:Work}).

\textbf{At minimal heat}. The first tri-control problem we will explore is to find the optimal protocols for the full parameter vector $\underline{\alpha}$ at minimal dissipation, with some consideration on whether the system ends in a steady-state or not. For trap parameters $\alpha_{1,2}$ we keep the end-points the same as was done for the simpler control problems ($\alpha_1 : 1 \rightarrow 5$, $\alpha_2 : 0 \rightarrow 1$), so as to facilitate comparison. Further, to the same end, we still treat $\alpha_3$ as a flexible parameter with the same fixed end-points $\alpha_3 : 1 \rightarrow 1$ and bounds $\alpha_3(t) \in (0.2,2]$. For completeness, all the objects defining our cost function for this scenario are given as:
\begin{align}
    \langle \mathcal{L}_W \rangle &= \frac{\dot{\alpha}_1(t)}{2}\left(m_{r^2}(t) - 2 \alpha_2(t) m_r(t) + \alpha_2^2(t)\right)  \label{eq:tc_LW}\\
    & \quad - \alpha_1 (t) \dot{\alpha}_2 (t) \left(m_r(t) - \alpha_2(t)\right),  \nonumber \\
    \langle \mathcal{L}_H \rangle &= \langle \mathcal{L}_W \rangle - \alpha_1(t) m_{rv}(t) + \frac{D'}{\mu \alpha_3(t)},\label{eq:tc_LH} \\
    B_H &= - \left[ \frac{\alpha_1(t)}{2} \left( m_{r^2}(t) - 2 \alpha_2(t) m_r(t) + \alpha_2^2(t) \right) \right]_0^{t_p},\label{eq:tc_BH} \\
    B_H &\underset{sts}{=} \frac{D'}{2\mu} \left( \frac{1}{1 + \mu \alpha_1(0) \alpha_3(0)} - \frac{1}{1 + \mu \alpha_1(t_p) \alpha_3(t_p)} \right) \label{eq:tc_BH_2}.
\end{align}

We show, for various $t_p$, the optimal tri-control protocols $\underline{\alpha}^\ast$ in Figs.~\ref{fig:tc_heat}A-D (no STS) and Figs.~\ref{fig:tc_heat}E-H (STS). Rather unexpectedly, given the complexity of the control, we find that the dissipation minimizing control protocols for each parameter are highly similar to those found in the simpler control problems. For the case of $\alpha_1$, at no STS transformation, the comparisons to make are Fig.~\ref{fig:tc_heat}A with Figs.~\ref{fig:casert}C, \ref{fig:dc_ol_stiffnesslocation_heat}A, and \ref{fig:dc_ol_stiffnesspersistence_heat}A; For the case of $\alpha_2$ (no STS) the comparisons to make are Fig.~\ref{fig:tc_heat}B with Fig.~\ref{fig:dc_ol_stiffnesslocation_heat}B; For $\alpha_3$ one should compare Fig.~\ref{fig:tc_heat}C with Fig.~\ref{fig:dc_ol_stiffnesspersistence_heat}B. The same conclusions can be drawn regarding the STS transformation optimal protocols, with an interesting (slight) deviation for the $\alpha_3$ protocol. Specifically, at a very late stage of the control we find a peculiar dip in the persistence, resulting in values lower than the default value of unity, (for all the $t_p$s explored here, see Fig.~\ref{fig:tc_heat}H). This essentially destroys the time-symmetry of the persistence protocols found earlier and is at odds with the no STS $\alpha_3^\ast(t)$, which was not seen in the dual control scenario (Figs.~\ref{fig:dc_ol_stiffnesspersistence_heat}B, F). Because this asymmetry is not seen in the no STS case the difference must come from the difference in boundary terms in the cost ($B_H$) (\ref{eq:tc_BH}, \ref{eq:tc_BH_2}). We attribute this late-stage asymmetry in the STS tri-control protocols to the dynamic coupling between the trap center, trap stiffness, and the active variance. By separating the spatial second moment into the centralized variance $V_r(t)$ and the squared mean, the stiffness contribution to the Lagrangian can be partitioned as $\frac{\dot{\alpha}_1}{2} V_r(t) + \frac{\dot{\alpha}_1}{2}(\alpha_2(t) - m_r(t))^2$. This reveals that stiffening the trap ($\dot{\alpha}_1 > 0$) incurs a thermodynamic penalty not only for compressing the particle's fluctuations, but also for any mean positional lag $(\alpha_2 - m_r)$ as the particle trails the moving trap center. As $t \to t_p$, the trap center is moving rapidly while the stiffness is peaking, causing this positional lag penalty to spike. Because the STS boundary term \eqref{eq:tc_BH_2} is fixed, the optimizer must absorb this cost in the bulk integral. To compensate for this unavoidable mechanical work, the optimizer aggressively suppresses the spatial variance $V_r(t)$ by plunging the persistence $\alpha_3(t)$ below unity, rapidly ``cooling'' the active fluctuations just before the protocol terminates. In the no STS case, the terminal internal energy is subtracted from the cost via \eqref{eq:tc_BH}, meaning the optimizer is not penalized for exiting the protocol with a large positional lag, entirely removing the need to kill the persistence. This also explains why this asymmetric dip is absent in the dual-control scenario (Fig.~\ref{fig:dc_ol_stiffnesspersistence_heat}F): with a static trap center ($\alpha_2 = 0$), the positional lag is strictly zero, and the variance suppression is not required. To help understand how (not) enforcing an STS effects some of the moments, we plot $\langle r(t) \rangle$ and $\langle rv(t) \rangle$ in Figs.~\ref{fig:tc_heat}I, J respectively. We observe higher $\langle r \rangle$ and lower $\langle rv \rangle$ for the STS transformation as compared to the no STS case.

\begin{figure}[t!]
    \centering
    \includegraphics[width=0.95\linewidth]{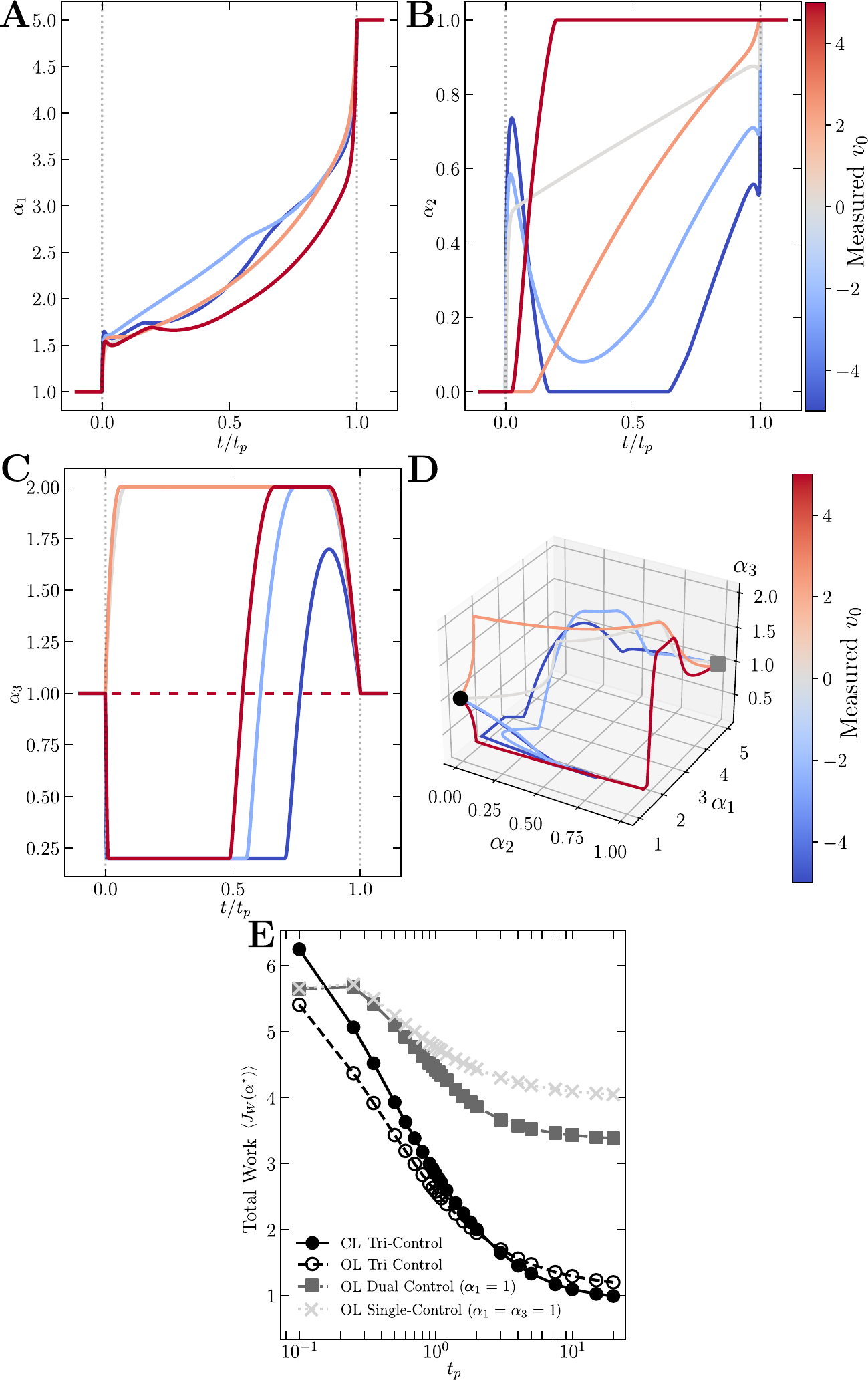}
    \caption{\textbf{Tri-control at minimal work.} \textbf{A} Optimal trap stiffness protocols as a function of $t/t_p$, for different $v_0$. \textbf{B} Optimal CL trap center protocols. \textbf{C} Optimal CL persistence protocols. \textbf{D} CL tri-control phase portrait. \textbf{E} The total work cost for the optimal tri-controls, as shown for four different cases (see legend). Model parameters: $\alpha_1(t=0^-) = 1$, $\alpha_1(t=t_p^+) = 5$, $\alpha_2(t=0^-) = 0$, $\alpha_2(t=t_p^+) = 1$, $\alpha_3(t=0^-) = \alpha_3(t=t_p^+) = 1$, $\alpha_1(t) \in (0,5]$, $\alpha_2(t) \in (0,1]$, $\alpha_3(t) \in (0.2,2]$, $D'=2$, $v_0 \in (-5,5)$, $t_p=1$, and $D=\mu=1$. Optimization parameters: $M=1000$ and $m_\varepsilon = 10^{-4}$. }
    \label{fig:tc_work}
\end{figure}

Lastly, we now analyse how the efficiency of the tri-control problem compares to that of the single- and dual- controls. There are many ways to make such comparisons, and we make the choice that we believe to be experimentally relevant: for the dual control we consider translocating an active particle with at constant activity, and for the single control we consider solely controlling the trap center. Thus, in all the comparisons the movement of the trap is the primary control goal. For this comparison, we find that having more control results in lower dissipation, with greater efficiency gains for longer protocols (see Figs.~\ref{fig:tc_heat}K,L). The greatest savings in heat occur at the optimal protocol durations, which is more clearly shown in the STS case (Fig.~\ref{fig:tc_heat}L), as was also observed in the dual control problem (see Fig.~\ref{fig:dc_ol_stiffnesspersistence_heat}D,H). 

\textbf{At minimal work}. Our final control problem involves deriving the minimal-work control vector $\underline{\alpha}^\ast$ with the privilege of an initial measurement of the self-propulsion $v_0$. There are two reasons motivating the construction of this problem: (i) the first is to understand whether the optimal tri-controls are similar to those obtained for the simpler control scenarios, as we witnessed for the OL tri-control at minimal heat, and (ii) the second is to see if efficiency gains can be observed for the work, as is the case for the dissipation. The work Lagrangian for this case is given as (\ref{eq:tc_LW}) with the replacements of the moments $m_X$, where $X=r,r^2$, with their conditional forms $m_X \vert_{v_0}$.

Overall, whilst most of the optimal protocols resemble (or build on) some of the qualitative features as was seen in the single- and dual-control scenarios, the optimal CL tri-control protocols are different (Figs.~\ref{fig:tc_work}A-D). For example, we find that $\alpha_1$ now exhibits noticeable, albeit small, oscillations (Fig.~\ref{fig:tc_work}A) that were absent in the dual control cases (Figs.~\ref{fig:dc_cl_stiffnesslocation_work}A and \ref{fig:dc_cl_stiffnesspersistence_work}A). The trap center $\alpha_2$ protocols also show significant deviations from the dual control case (compare Figs.~\ref{fig:tc_work}A with \ref{fig:dc_cl_stiffnesslocation_work}B), which includes an observable--though small--dip in the trap center before the end of the protocol. The protocols for the persistence parameter show the greatest differences to the dual control scenarios (compare Fig.~\ref{fig:tc_work}C with Figs.~\ref{fig:dc_cl_stiffnesspersistence_work}B and \ref{fig:dc_ol_stiffnesspersistence_work}B). The main difference is observed for high self-propulsion measurements ($|v_0| \gtrapprox 2$), where $\alpha_3$ is first rapidly decreased to the lower bound and then at, or just past, $t/t_p \approx 1/2$ the persistence parameter is increased to the upper bound and is, finally, brought back to $\alpha_3 = 1$. We stress this is not just quantitatively different to the simpler control scenarios, it is qualitatively different and only arises due to the interplay with the other two control parameters $\alpha_{1,2}$. Finally, we compare the total work for the CL tri-control problem to other--more simple--problems (Figs.~\ref{fig:tc_work}E,F). As with the OL tri-control at minimal heat, the primary aim is to move the trap and so $\alpha_2$ is never fixed. Then, for the simpler control problems we choose OL tri-control, to examine the role of measurement, OL dual control at fixed $\alpha_1$, as we have yet to compare controls that exploit flexibility of $\alpha_3$, and OL single control at fixed $\alpha_{1,3}$, where we choose open-loop as opposed to closed-loop to compare to the most basic translocation control. Importantly, we find that for durations $t_p \gtrapprox 2-3$, having both more control, \emph{i.e.,} more degrees of freedom, and more information (initial measurement) results in the most efficient control (Fig.~\ref{fig:tc_work}E). However, for the fastest controls explored here $t_p = 0.1$, CL tri control is the least efficient, with OL tri control being the most optimal. This is, at least partly, explained by the same physics behind the inset of Fig.~\ref{fig:dc_cl_stiffnesslocation_work}D: the conditioned work becomes convex in $v_0$ with greater costs for higher $|v_0|$ at short $t_p$. 

\textbf{Surprising effectiveness of ``naively'' superposing single controls for multi-parameter control}. Motivated by practical considerations, we next wondered how much worse is it if one ``naively'' takes the optimal single-control protocols and uses them simultaneously for the tri-control? As expected, we find that this naive approach to multi-parameter control yields higher thermodynamic costs compared to the full optimized tri-controls (see Fig.~\ref{fig:superposition}). However, rather surprisingly, we find that the extra work on the controller is within a conservative $\sim 5-10\%$ of the fully optimized work (see Figs.~\ref{fig:superposition}A,B). This finding is robust for minimizing the work in both dual- and tri-control scenarios. Interestingly, we found that this is also true for the total heat, though we do observe greater efficiency gains for the fully optimized solutions for closed-loop control Figs.~\ref{fig:superposition}C,D. We note that the closed-loop protocols yield negative total heat (cooling) for certain protocol durations, with the cooling effect being more pronounced in the no STS case. On first inspection this seems rather counter-intuitive, however this does not violate the second law \cite{Datta2022,Liu2026}; rather, it reflects the well-known paradigm of information thermodynamics where the controller acts as a Maxwell's demon \cite{Cocconi2024,GarciaMillan2025,Schuttler2025}. Because our thermodynamic cost functional does not account for the memory erasure costs associated with measuring the initial self-propulsion, the controller is allowed to extract work from the active bath, resulting in net cooling of the system.

\section{Discussion}
\label{sec:6}

\begin{figure}[t!]
    \centering
    \includegraphics[width=0.95\linewidth]{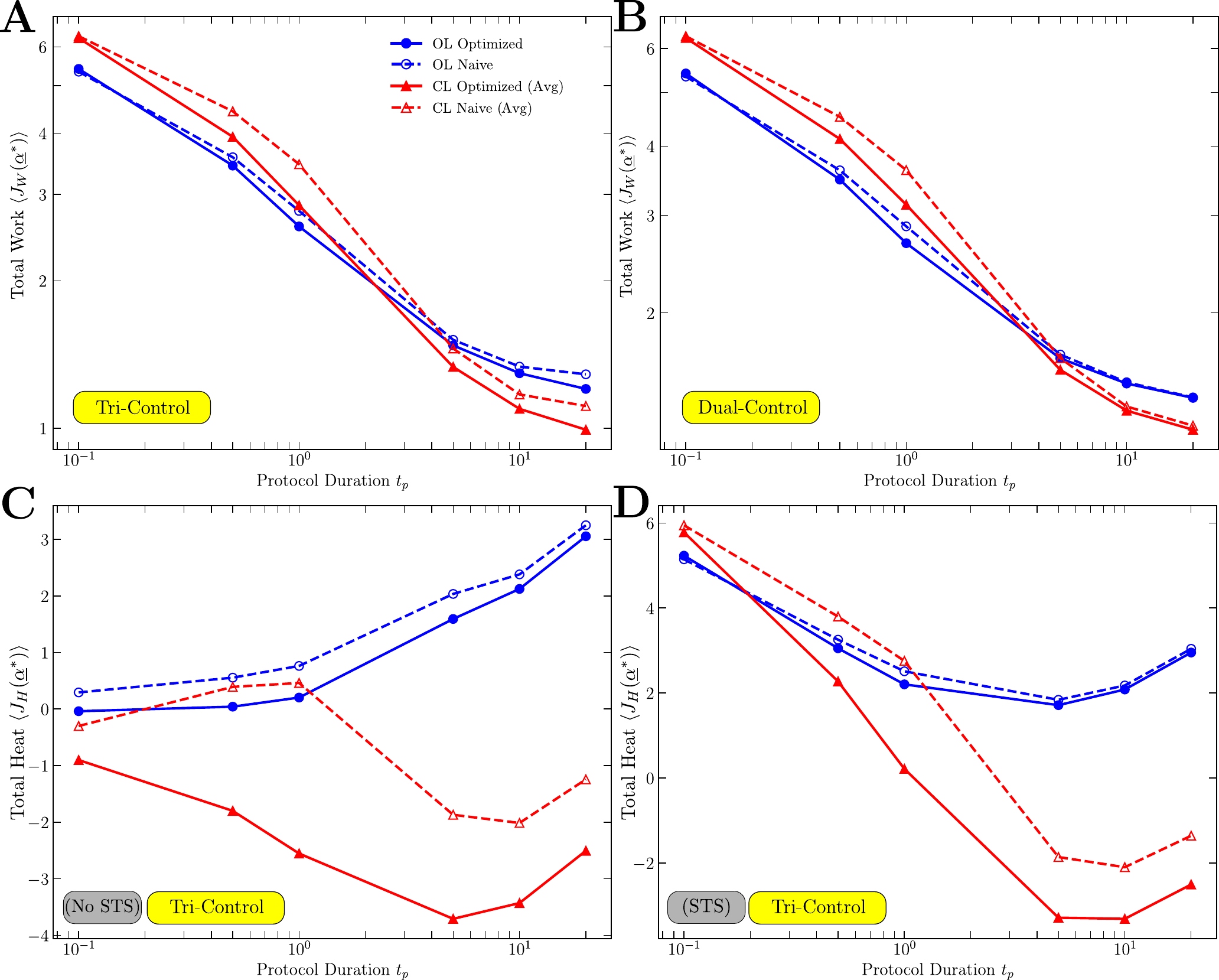}
    \caption{\textbf{Comparing the simultaneous driving of independently optimized controls with fully optimized multi-parameter controls.} \textbf{A} Comparing the total work as a function of $t_p$ for fully optimized tri controls (solid symbols) and for the simultaneous driving of independently optimized controls (open symbols). We show both open-loop (OL, blue) and closed-loop (CL, red) solutions. Only a slightly higher cost is observed for naively driving single-parameters in unison. \textbf{B} Same as (A), but for dual control at fixed activity $\alpha_3 =1$. \textbf{C} Same as (A) but now comparing the total heat (no STS). For the closed-loop tri control we observe greater gains in efficiency for the fully-optimized solutions, and we also observe cooling of the bath (see main text). \textbf{D} Same as (C) but for an enforced STS transformation. The same model and optimization parameters were used as in Figs.~\ref{fig:dc_ol_stiffnesslocation_work} (for B), \ref{fig:dc_cl_stiffnesslocation_work} (for B), \ref{fig:tc_work} (A and B), and \ref{fig:tc_heat} (for C and D) , for the matching control regime.}
    \label{fig:superposition}
\end{figure}

The realization of small-scale efficient control of active matter is currently challenging, in part due to a growing gap between what is achievable in theory and what is attainable in experiment. While experimentalists can simultaneously control optical trap strengths and locations \cite{Blaber2022}, with the potential for activity control \cite{Truong2026}, theoretical solutions that account for such microscopic controls have largely remained confined to single-parameter driving, or asymptotic time limits \cite{Davis2024}. In this work, we have taken a step toward bridging this gap by developing a robust computational framework to solve fully coupled, multi-parameter active control problems, free from any protocol duration assumptions. By leveraging gradient descent and exact reverse-mode automatic differentiation, which has seen previous success in the finite-time driving of equilibrium systems \cite{Engel2023}, our approach provides a transparent alternative to more ``black-box'' machine-learning methods, such as neural evolution \cite{Casert2024}.

To establish the validity of our framework, we tested it against exact analytical solutions for single-parameter control: the classic Schmiedl--Seifert control problem \cite{Schmiedl2007}, and recent results for closed-loop active trap translation \cite{Schuttler2025}. A central mathematical feature of optimal thermodynamic control is that the cost functionals---both work and heat---are affine in the control speeds. By Pontryagin's maximum principle, this affine structure implies that the exact mathematical optima are discontinuous ``bang-bang'' protocols, which manifest numerically as infinite-frequency chattering (known as Fuller's phenomenon) \cite{Fuller1963}. We demonstrated that introducing a quadratic kinetic cost for the controls effectively regularizes the optimization, removing discontinuities and yielding smooth, experimentally realizable protocols. While the physical justification for this smoothness (experimental bandwidth limits and finite actuation costs) differs from the smoothness naturally emerging in linear-response-based frameworks \cite{Sivak2012,Davis2024}, both share a mathematical origin in breaking the strictly affine structure of the cost Lagrangian. Remarkably, we found that these smooth, regularized single-parameter protocols are highly efficient, yielding thermodynamic costs very close to those of the exact discontinuous solutions.

This high-fidelity regularization allowed us to uncover fine structural details in the optimal protocols. By overlaying the optimal closed-loop trap-center protocols conditioned on various initial self-propulsions $ v_0 $, we identified a distinct ``piranha''-shaped structural motif---an initial fanning out of the trap trajectories, a finite-time re-intersection, and a subsequent secondary fanning. Crucially, we found that this structural signature is not limited to isolated trap translation, but persists in the fully coupled multi-parameter control scenarios studied here. Furthermore, when comparing our method to protocols derived via neural-network evolution \cite{Casert2024}, we found broad agreement. However, the exact-gradient precision of our AD framework revealed a subtle, late-stage non-monotonic ``breathing'' mode in the trap stiffness that was not resolved by that method, highlighting the utility of transparent computational methods in exploring optimal control strategies.

Scaling the framework to multi-parameter trap scenarios revealed a rich landscape of active thermodynamic control. In the dual-control regime, manipulating the trap stiffness and persistence time unveiled a clear divergence between work-minimization and heat-minimization strategies. Because the work cost is dictated by mechanical compression, thereby imposing a distinct ``arrow of time,'' the optimal activity protocol is highly asymmetric, delaying the reduction of active fluctuations until the late stages of the protocol to suppress variance penalties. Conversely, the heat-optimal protocols are symmetric, reflecting the need to suppress the state-independent housekeeping dissipation throughout the protocol. Extending this to the tri-control regime---simultaneously tuning stiffness, center, and persistence---revealed nontrivial cross-couplings. Most notably, when enforcing a strict state-to-state (STS) transformation, the optimizer induces a sharp, asymmetric late-stage dip in the persistence. This feature emerges to dampen the active fluctuations.

Overall, for manipulation of trapped particles, we find that optimizing multi-parameter controls results in lower thermodynamic work and dissipation. This finding is consistent with the dual control problem studied in \cite{Blaber2022}. One of the most practically significant findings of this work is the emergence of a near-superposition principle for multi-parameter active control, which we systematically verified across both work and heat minimization strategies (see Fig.~\ref{fig:superposition}). We demonstrated that naively executing independently optimized single-parameter protocols simultaneously incurs only roughly an extra 5--10\% of total work as compared to the fully coupled multi-parameter solution. As shown in Figs.~\ref{fig:superposition}A and B, this weak sub-optimality holds robustly for the mechanical work across all protocol durations in both dual- and tri-control settings. Interestingly, while this principle also holds broadly for heat dissipation (Figs.~\ref{fig:superposition}C and D), we observe slightly larger efficiency gains for the fully coupled solutions in the closed-loop regime. We note that this finding resonates with previous work that also explored the practical considerations of simple control strategies, and found that fully optimized solutions always outperform naive ones \cite{Zulkowski2012,Zulkowski2013,Wareham2025}. However, we note that what we mean by naive, running independently \textit{optimized} controls, is different to what is typically considered naive: executing linear protocols with \textit{no optimization}.

Looking forward, this computational framework opens several promising avenues for non-equilibrium (active) control. A natural future direction is the multi-parameter optimal control of many-body active matter, where inter-particle interactions could generate substantially more complex optimal protocols. A key challenge there will be to carefully account for stochastic trajectories (see also \cite{Engel2023}). Further application of our method could also facilitate the exploration of the control of active field theories, as an approach to go beyond the restrictions of linear response \cite{Soriani2025}. We also wonder whether connections can be made between the kinetically costed, regularized protocols found here and the geodesic-counterdiabatic approach used in \cite{Zhong2024}. Finally, the optimal driving of active dynamical phase transitions, such as MIPS, while still a formidable challenge, is at least now more accessible using optimal multi-parameter control frameworks similar to the one developed here.

\begin{acknowledgments}
     L.K.D. acknowledges the interesting discussions had with Alessandro Manacorda, \'{E}tienne Fodor, Karel Proesmans, John Bechhoefer, David Sivak, Sabine Klapp, and Deepak Gupta. L.K.D. acknowledges the Flora Philip Fellowship at the University of Edinburgh and also acknowledges funding from the Isaac Newton Institute for Mathematical Sciences Postdoctoral Research Fellowship (EPSRC Grant Number EP/V521929/1).
\end{acknowledgments}

\appendix
\beginsupplement

\section{Protocol jumps for thermodynamic costs}
\label{app:Jumps}
This can be formalized by converting the protocol optimization into a standard bounded-control problem. Introduce the control speed $\underline{u}(t) := \dot{\underline{\alpha}}(t)$ and treat $\underline{\alpha}$ as an additional state with dynamics $\dot{\underline{\alpha}} = \underline{u}$, with admissible speeds $\underline{u}(t) \in \mathcal{U} := \prod_{a=1}^n [u_a^{\min}, u_a^{\max}]$. Both objectives take the generic form:
\begin{equation}
\label{eq:affine_generic_cost}
    J = B + \int_0^{t_p} dt \, \Big[ g(\underline{X}, \underline{\alpha}, t) + \underline{u}^\intercal(t) \underline{\lambda}(\underline{X}, \underline{\alpha}, t) \Big],
\end{equation}
with $g$ independent of $\underline{u}$. The PMP Hamiltonian for the augmented system is:
\begin{equation}
\label{eq:PMP_Hamiltonian}
    \mathcal{H} = \underline{p}_X^\intercal f(\underline{X}, \underline{\alpha}, t) + g(\underline{X}, \underline{\alpha}, t) + \underline{u}^\intercal \big( \underline{p}_\alpha + \underline{\lambda} \big),
\end{equation}
so the control enters affinely. The pointwise minimizer therefore lies on the boundary of $\mathcal{U}$:
\begin{equation}
\label{eq:bangbang_PMP}
\begin{aligned}
    \underline{u}^\ast(t) &\in \arg\min_{\underline{u} \in \mathcal{U}} \underline{u}^\intercal \underline{\sigma}(t), \quad \underline{\sigma}(t) := \underline{p}_\alpha(t) + \underline{\lambda}(t).
\end{aligned}
\end{equation}
For the box constraint (whenever $\sigma_a(t) \neq 0$) one has the bang–bang law:
\begin{equation}
\label{eq:bang_bang_control}
    u_a^\ast(t) = 
    \begin{cases} 
        u_a^{\min}, & \sigma_a(t) > 0, \\
        u_a^{\max}, & \sigma_a(t) < 0,
    \end{cases}
\end{equation}
with switches at the (typically isolated) times where $\sigma_a(t) = 0$. In the large-actuation limit $|u_a^{\max}| \to \infty$, these saturated segments shrink to impulses, recovering jump discontinuities in $\alpha_a(t)$.

\section{Fuller's phenomenon in solving optimal thermodynamic control}
\label{app:Chattering}
For $n > 1$ one generically encounters times where several components of the switching vector $\underline{\sigma}(t)$ are small, and the Hamiltonian minimization \eqref{eq:bangbang_PMP} becomes non-unique. A (multi-)singular arc corresponds to an interval on which, for some index set $I \subset \{1, \ldots, n\}$,
\begin{equation}
\label{eq:singular_arc_condition}
    \sigma_a(t) \equiv 0 \qquad \forall a \in I,
\end{equation}
so that first-order optimality leaves the corresponding components of $\underline{u}$ undetermined (any value in the relevant face of $\mathcal{U}$ gives the same $\mathcal{H}$). To determine a singular control one differentiates \eqref{eq:singular_arc_condition} until the control appears explicitly:
\begin{equation}
\label{eq:singular_higher_order}
\begin{aligned}
    0 &= \frac{d^q}{dt^q} \sigma_a(t) \Big|_{\text{sing.}} \\
    &= \Phi_a^{(q)} \big( \underline{X}(t), \underline{\alpha}(t), \underline{p}(t), \underline{u}(t) \big), \qquad a \in I,
\end{aligned}
\end{equation}
with $q$ the smallest order such that $\partial \Phi_a^{(q)} / \partial \underline{u} \neq 0$. The resulting $\underline{u}_{\text{s}}(t)$ typically lies in the interior of the admissible polytope $\mathcal{U}$. However, since $\mathcal{H}$ is linear in $\underline{u}$, interior values are not selected by pointwise minimization; instead the optimal admissible control is obtained as a limit of ever faster switches between vertices of $\mathcal{U}$ whose convex combination realizes $\underline{u}_{\text{s}}(t)$. 

At junctions between bang arcs and a higher-order singular arc (order $q \ge 2$), the switching times accumulate geometrically at a finite time $t_F$,
$    t_k \uparrow t_F, t_F - t_{k+1} \sim \rho (t_F - t_k)$, for $0 < \rho < 1,$ producing infinitely many switches in finite time: Fuller's phenomenon. Physically, multi-parameter thermodynamic controllers then ``dither'' between extremal protocol velocities, i.e., they chatter between parameter-space directions/endpoints, in order to effectively implement the interior singular control dictated by the coupled (state-dependent) coefficients $\underline{\lambda}(t)$.

\section{Gradient descent via automatic differentiation}
\label{app:GDAD}

To obtain noise-averaged, or conditioned, optimal protocols we first parameterize the protocol for control parameter $l$ by $M$ uniformly spaced time nodes on the interval $[0,t_p]$ as:
\begin{equation}
\begin{aligned}
        &\underline{a}^{(l)} = (a_1^{(l)},\ldots,a_M^{(l)})^\intercal \in \mathbb{R}^M, \\
        &s_j = \frac{j-1}{M-1}t_p, \quad j=1,\dots, M,
\end{aligned}
\end{equation}
with $\alpha_l(s_j) = a_j^{(l)}$, where the continuous control ${\alpha}_l(t)$ is gotten from a piecewise-linear interpolation and implemented as a differentiable map ${\alpha}_l(t) = \mathcal{I}[\underline{a}^{(l)}](t)$.

Vanilla gradient descent involves the following:
\begin{equation}
\underline{a}^{(n+1)} = \underline{a}^{(n)} - \eta \nabla_{\underline{a}}\langle J \rangle (\underline{a}^{(n)}),
\end{equation}
with learning rate $\eta>0$. In practice we employ the \texttt{Adam} optimizer, which maintains exponentially moving averages of the first and second moments of the gradient \cite{Kingma2014}. At iteration $t$ with gradient $\underline{g}_t=\nabla_{\underline{a}}\langle J \rangle (\underline{a}_t)$, \texttt{Adam} updates are
\begin{equation}
\begin{aligned}
\underline{m}_t &= \beta_1 \underline{m}_{t-1} + (1-\beta_1) \underline{g}_t, \\
\underline{v}_t &= \beta_2 \underline{v}_{t-1} + (1-\beta_2) \underline{g}_t^{\circ 2},\\
\hat{\underline{m}}_t &= \frac{\underline{m}_t}{1-\beta_1^t}, \qquad \hat{\underline{v}}_t = \frac{\underline{v}_t}{1-\beta_2^t}, \\
\underline{a}_{t+1} &= \underline{a}_t - \eta \frac{\hat{\underline{m}}_t}{\sqrt{\hat{\underline{v}}_t}+\varepsilon'},
\end{aligned}
\end{equation}
where $\beta_1\in(0,1)$, $\beta_2\in(0,1)$, $\varepsilon'>0$ is a small stabilizer, and $\circ 2$ denotes elementwise squaring.

We use \texttt{JAX} \cite{jax2018github} to compute the explicit gradients $\nabla_{\underline{a}}\langle J \rangle$ via reverse-mode automatic differentiation. This is greatly facilitated by the simplicity of the model (see below), considering a single trapped particle in one-dimension, with the numerical simulation model involving only the time integration of deterministic ODEs for the relevant moments.

\begin{figure*}[h!]
    \centering
    \includegraphics[width=0.95\linewidth]{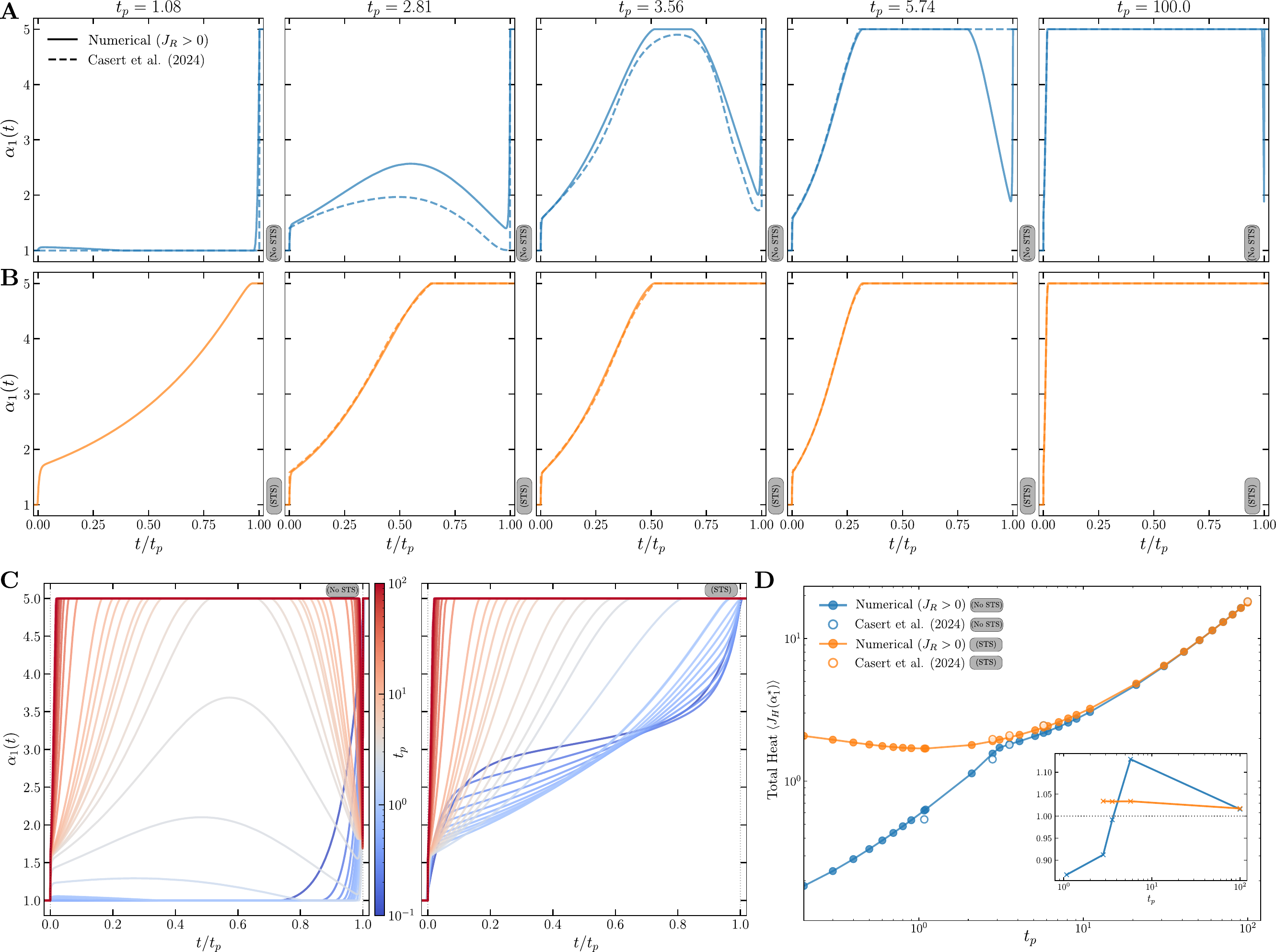}
    \caption{ \textbf{Comparing our single control results to that derived from neural evolution (Casert et. al. 2024) \cite{Casert2024}}. \textbf{A} Optimal protocols for the trap stiffness $\alpha_1(t)$ as a function of $t/t_p$, for five selected $t_p$ values, as found using the exact-gradient AD optimization used here (numerical, solid lines) and as found in \cite{Casert2024}. The optimization is done without imposing a state-to-state transformation (no STS). Note that the neural evolution protocols miss the late stage ``breathing'' of the trap for slower durations. \textbf{B} The same as (A) but for an enforced STS transformation. \textbf{C} Optimal trap stiffness protocols for all the $t_p$ values explored here. Again, notice the non-monotonicity of the protocols for $t_p \gtrapprox 2$. \textbf{D} Total heat dissipation for the optimal stiffness protocols shown for both regularized AD (solid circles) and neural evolution (open circles). Inset shows the neural evolution heat values divided by the AD values. Model parameters: $\alpha_1(t=0^-) = 1$, $\alpha_1(t=t_p^+) = 5$, $\alpha_2=0$, $\alpha_3 =1$, $D'=2$. Optimization parameters: $M=500$, $m_\varepsilon = 10^{-4}$. Neural evolution heat values were computed from protocol data made available in \cite{Casert2024}.}
    \label{fig:casert}
\end{figure*}

\bibliography{refs_new}

\end{document}